\documentclass[fleqn,usenatbib]{mnras}
\usepackage[T1]{fontenc}
\usepackage{ae,aecompl}
\usepackage{graphicx}
\usepackage{amsmath}
\usepackage{amssymb}
\usepackage{txfonts}

\title[Splitting the lentils]{Splitting the lentils: Clues to galaxy/black hole coevolution
  from the discovery of offset relations for non-dusty versus dusty (wet-merger-built) lenticular
  galaxies in the $M_{\rm bh}$-$M_{\rm *,spheroid}$ and $M_{\rm bh}$-$M_{\rm
    *,galaxy}$ diagrams}

\author[Graham]
{
Alister W.\ Graham$^{1,2}$\thanks{E-mail: AGraham@swin.edu.au}
\\
$^1$ Centre for Astrophysics and Supercomputing, Swinburne University of
Technology, Hawthorn, VIC 3122, Australia
\\
$^2$ OzGrav-Swinburne, Centre for Astrophysics and Supercomputing, Swinburne
University of Technology, Hawthorn, VIC 3122, Australia
}

\date{Accepted 15 Feb 2023. in original form 31 Dec 2022.}
\pubyear{2023}

\begin{document}
\label{firstpage}
\pagerange{\pageref{firstpage}--\pageref{lastpage}}
\maketitle

\begin{abstract}

This work advances the (galaxy morphology)-dependent (black hole mass, $M_{\rm
  bh}$)-(spheroid/galaxy stellar mass, $M_*$) scaling relations by introducing
`dust bins' for lenticular (S0) galaxies.  Doing so has led to the discovery of
$M_{\rm bh}$-$M_{\rm *,sph}$ and $M_{\rm bh}$-$M_{\rm *,gal}$ relations for
dusty S0 galaxies---built by major wet mergers and comprising half the S0
sample---offset from the distribution of dust-poor S0 galaxies.  The
situation is reminiscent of how  major dry mergers of massive S0 galaxies have
created an offset population of ellicular and elliptical galaxies.  For a
given $M_{\rm bh}$, the dust-rich S0 galaxies have 3--4 times higher $M_{\rm
  *,sph}$ than the dust-poor S0 galaxies, and the steep distributions of both
populations in the $M_{\rm bh}$-$M_{\rm *,sph}$ diagram bracket the $M_{\rm
  bh} \propto M_{\rm *,sph}^{2.27+/-0.48}$ relation defined by the spiral
galaxies, themselves renovated through minor mergers.  The new relations offer
refined means to estimate $M_{\rm bh}$ in other galaxies and should aid with: 
(i) constructing (galaxy morphology)-dependent black hole mass functions; 
(ii) estimating the masses of black holes associated with tidal disruption
events; 
(iii) better quantifying evolution in the scaling relations via improved comparisons
with high-$z$ data by alleviating the pickle of apples versus oranges; 
(iv) mergers and long-wavelength gravitational wave science;
(v) simulations of galaxy/black hole coevolution 
and semi-analytic works involving galaxy speciation; plus 
(vi) facilitating improved
extrapolations into the intermediate-mass black hole landscape.  The role of
the galaxy's environment is also discussed, and many potential projects that can
further explore the morphological divisions are mentioned.

\end{abstract}


\section{Introduction} 
\label{Sec_Intro}

Stars create metals and dust, polluting or rather fertilising galaxies.
Today, there is typically a couple of hundred times more total-gas mass than
dust mass, at least in local late-type galaxies (LTGs) where the mass ratio
ranges from 10 to 1000, and $1.5 \lesssim \log(M_{\rm H{\footnotesize
    I}}/M_{\rm dust}) \lesssim 3$ dex \citep[][their Fig.~2 and 3, and
  Table~4]{2020A&A...633A.100C, 2022A&A...668A.130C}.  In an ensemble of
early-type galaxies (ETGs) detected by the Herschel Observatory, the average
dust-to-stellar mass ratio has been reported as $\log(M_{\rm dust}/M_*) =
-4.3$ dex by \citet{2012ApJ...748..123S}.  Despite the low fraction of dust,
when not diffusely distributed \citep[e.g.][]{1989ApJS...70..329K,
  1995A&A...298..784G}, it can be easy to spot in silhouette against the
uniform screen of an ETG's optical light \citep[e.g.][]{1985MNRAS.214..177S,
  1988A&A...204...28V}. It can appear as expansive dust lanes
patchy clouds, and tangled serpentine shapes, formed in dense molecular
clouds, where the dust has
condensed into larger particles with (typically CO and N) ice mantles and
polycyclic aromatic hydrocarbons, built in part from the larger reservoir of
dust, carbon and silicate grains, and atomic and ionic metals in the diffuse
interstellar medium.

The ETGs can be built through collisions.  Violent mergers which erase the orbital
angular momentum of the progenitor system form elliptical (E) galaxies.  When
the bulk of the orbital angular momentum is not cancelled, major collisions
will produce lenticular (S0)\footnote{Previously thought to arise before Sa
  galaxies, the S0 galaxy type was theorised by \citet{1919pcsd.book.....J}
  and identified by \citet[][see his p.1016]{1925MNRAS..85.1014R}.} disc
galaxies \citep[e.g.][]{2003ApJ...597..893N}, such as NGC~5128, aka
Centaurus~A, with a rotating structure and a `bulge' component
\citep{1990AJ.....99.1781V, 2016ApJS..222...10S}.
\citet{Graham:Sahu:22a, Graham:Sahu:22b} revealed how major (dry and
wet)\footnote{Wet mergers involve galaxies containing gas and star formation,
  while dry mergers are (cold gas)-poor mergers without star formation.}
galaxy mergers result in a `punctuated equilibrium' that does not maintain a
universal $M_{\rm bh}$-$M_{\rm *,gal}$, $M_{\rm bh}$-$M_{\rm *,sph}$, nor
$M_{\rm bh}$-$R_{\rm e,sph}$ scaling relation but instead yields a dramatic
jump when galaxies evolve from one morphological type to
another.\footnote{\citet{Graham-sigma} explains why major dry mergers do not
  yield a significant jump in the $M_{\rm bh}$-$\sigma$ diagram.}  Here, this is
explored more deeply by investigating the S0 galaxies, previously treated as a
single morphological type.  The exception to this statement is that four dusty
S0 galaxies had previously been flagged (and excluded) due to their known
merger-remnant status.

Motivating the current investigation were two observations/questions, which
turned out to be related.  The first was, Why do the spiral galaxies display a
steep $M_{\rm bh}$-$M_{\rm *,sph}$ relation splicing through the distribution
of S0 galaxies? This trend was seen in \citet{2019ApJ...876..155S} and
\citet{Graham:Sahu:22a} but needed to be explained.  The second was, Why do
the S0 galaxies display a large scatter in the $M_{\rm bh}$-$M_{\rm *,sph}$
diagram compared to the S and E galaxies?  This latter question, perhaps first
raised in \citet{Graham:Sahu:22b}, was the initial question tackled when
commencing this project, but it led to additionally solving the first mystery.
As shall be seen, the answers came about by first recognising that the
previously excluded S0 mergers contained copious amounts of dust and then
checking on the dust content and merger status of the other S0 galaxies with
directly measured black hole masses.  This endeavour has led to yet further
evidence that mergers, rather than just feedback from active galactic nuclei
(AGNs), are shaping the distribution of galaxies in the $M_{\rm bh}$-$M_{\rm
  *,sph}$ and $M_{\rm bh}$-$M_{\rm *,gal}$ diagrams and thus shaping their
evolution.  It is shown herein that dusty S0 galaxies built from wet mergers
tend to be offset to notably higher stellar masses and lower $M_{\rm
  bh}/M_{\rm *,sph}$ ratios than non-dusty S0 galaxies.  This offset can arise
from a partial redistribution of what were initially disc stars into the
merger remnant's bulge and ties in with the population of star-forming S0
galaxies, which tend to have centrally-dominated, rather than disc-dominated,
star formation \citep{2022MNRAS.513..389R}.

The ongoing discovery of (galaxy morphology)-dependent substructures in the
(black hole)-(stellar mass) diagrams represents significant advances for many
reasons.  Perhaps foremost is the understanding it brings to, and the
witnessing of, the hierarchical evolution of galaxies, together with a deeper
understanding of the range of $M_{\rm bh}/M_{\rm *,sph}$ ratios in galaxies.
Furthermore, awareness of a galaxy's dust content and structural morphology
can facilitate an improved prediction of its central black hole mass. For
example, for a given $M_{\rm bh}$, visibly dusty S0 galaxies tend to have
$M_{\rm *,sph}$ values some 3 to 4 times higher than in S0 galaxies without
visible signs of dust.  Improved estimates of black hole mass will aid a
plethora of related science, such as studies of tidal disruption events
\citep[e.g.][]{2015JHEAp...7..148K}, black hole accretion efficiencies
\citep[e.g.][]{2012MNRAS.419.2529R}, and gravitational wave science from
colliding massive black holes \citep[e.g.][]{2006PhRvD..73f4030B} and extreme
mass ratio inspiral events \citep{2009CQGra..26i4034G, 2012A&A...542A.102M}
fuelled by the coexistence of massive black holes and nuclear star clusters
\citep[e.g.][and references therein]{2020MNRAS.492.3263G}.

The following section introduces the data used for this study, which revolves
around 40 S0 galaxies (Table~\ref{Table-data}).  
Comparisons are also made with other galaxy types, and
the larger sample contains $\sim$100 galaxies with directly measured black
hole masses.  A handful of exclusions are described in Section~\ref{Sec_Excl}, 
also see Section~\ref{Sec_noncon}, along with a recognition that some of the
apparent outlying systems may hold clues to an even more intricate
understanding of the evolution of galaxies.  Given the use of
bulge stellar masses obtained from bulge luminosities (Section~\ref{Sec_MonL}),
Section~\ref{Sec_bulge} offers some additional insight into how bulges are
identified. It also discusses an emerging trend in which bulges are giving up
ground to other galaxy components, in a few cases leading to ambiguity over
whether a galaxy may be better considered bulgeless.  Circumventing potential 
dependence on the galaxy decompositions, the $M_{\rm bh}$-$M_{\rm *,gal}$
diagram is additionally used in Section~\ref{Sec_S0_split}, along with the
$M_{\rm bh}$-$M_{\rm *,sph}$ and $M_{\rm bh}$-$R_{\rm e,sph}$ diagrams, to
reveal the division between dusty and non-dusty S0 galaxies.  

The literature was searched for evidence of mergers and accretion in the 40 S0
galaxies, and the results are reported throughout and summarised in
Table~\ref{Table-dust}.  Recognising that the dusty S0 galaxies are the
product of major wet mergers, and given their location on the high $M_{\rm
  *,sph}/M_{\rm bh}$ side of the distribution of disc galaxies in the $M_{\rm
  bh}$-$M_{\rm *,sph}$ diagram, Section~\ref{Sec_Spiral} briefly reviews the
location of the S galaxies positioned between the visibly dust-rich and
dust-poor S0 galaxies.

Section~\ref{Sec_noncon} further reports on systems, i.e.,
spheroid/galaxy/black hole pairings, which do not conform to the general trends for
one reason or another; and may be worthy of further
consideration in future work.  
Having derived $M_{\rm bh}$-$M_{\rm *,sph}$ relations for dusty and
non-dusty S0 galaxies in Section~\ref{Sec_S0_split}, the latter is used in
Section~\ref{Sec_BonT} to 
derive the typical bulge-to-total ($B/T$) stellar mass ratio associated with a (dry
merger)-induced jump from the non-dusty S0 galaxy $M_{\rm bh}$-$M_{\rm *,sph}$
relation to the E galaxy $M_{\rm bh}$-$M_{\rm *,sph}$ relation, refining the
calculation in Graham (2022) that was based on a single $M_{\rm
  bh}$-$M_{\rm *,sph}$ relation for all S0 galaxies.  A discussion of several
topics  is provided in 
Section~\ref{Sec_Disc}, including the concept that some S0 galaxies are faded
S (and faded S0) galaxies, the role of environment, AGN feedback, the benefits
of refined black hole mass estimates, and more. 
Rather than reiterating the new relations for the S0 galaxies, some of the many
impacts and potential future projects they enable are listed in
Section~\ref{Sec_Future}.

Finally, 
one likely ingredient for understanding what is occurring may sound 
obvious, but the presence of gas reveals an environment which permits its
existence.  The dusty S0 galaxies have grown with neither
gas-stripping nor a central `Benson Burner'\footnote{Coined in footnote 33
  of \citet{Graham:Sahu:22a}, a `Benson Burner' refers to an AGN capable of
  maintaining a hot X-ray halo, as suggested by \citet{2003ApJ...599...38B}, 
  and implemented and observed in, for example, \citet{2006MNRAS.370.1651C} 
 and \citet{2007ARA&A..45..117M}.} curtailing gas-cooling and preventing star
formation. 
Therefore, a cursory investigation into the (cluster, group, or field) environment
has been made here.  While this did not prove fruitful, it may
be a helpful reference for future work and is included in an Appendix.

\begin{figure*}
\begin{center}
$
\begin{array}{ccc} 
\includegraphics[trim=0.0cm 0cm 0.0cm 0cm, height=0.3\textwidth, angle=0]{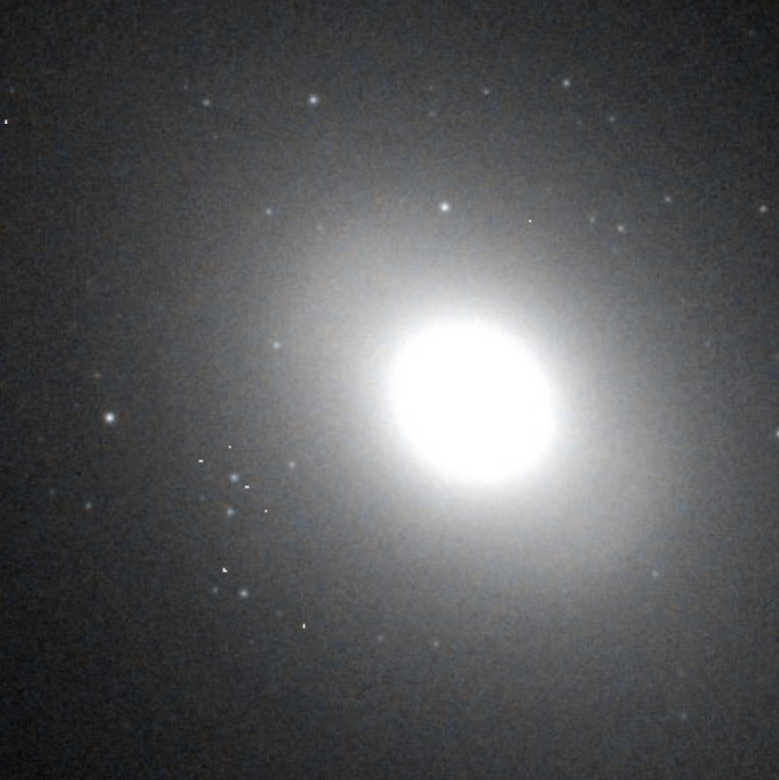} & 
\includegraphics[trim=0.0cm 0cm 0.0cm 0cm, height=0.3\textwidth, angle=0]{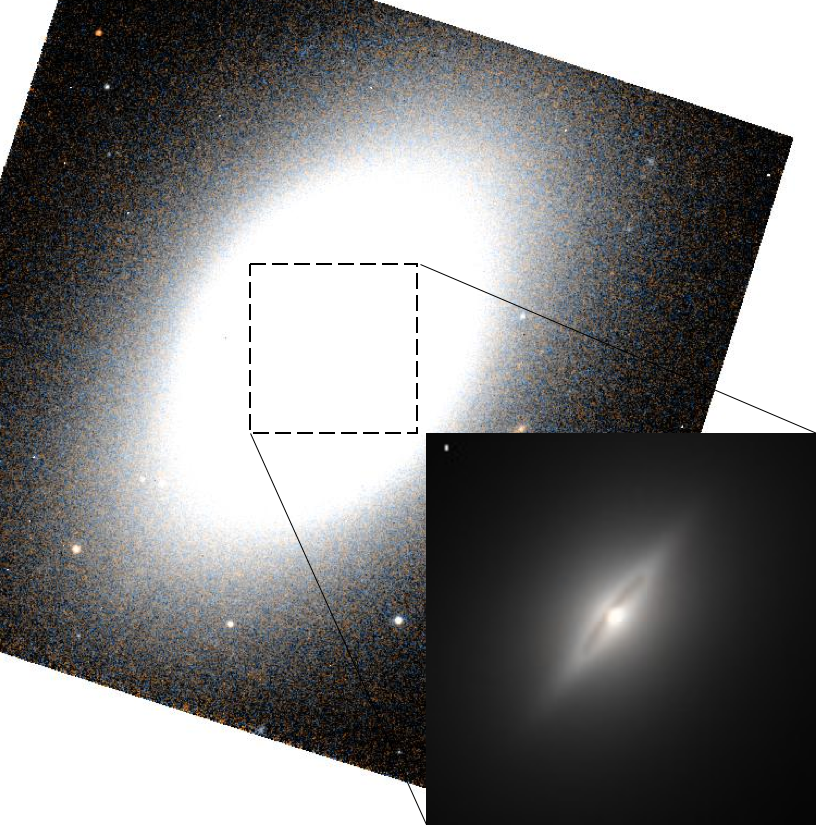} &
\includegraphics[trim=0.0cm 0cm 0.0cm 0cm, height=0.3\textwidth, angle=0]{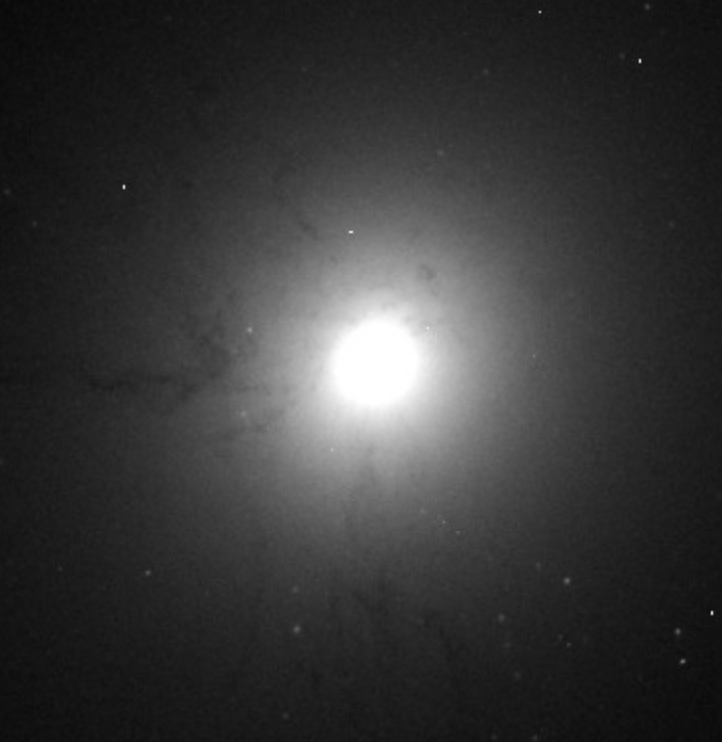} \\
\includegraphics[trim=0.0cm 0cm 0.0cm 0cm, height=0.3\textwidth, angle=0]{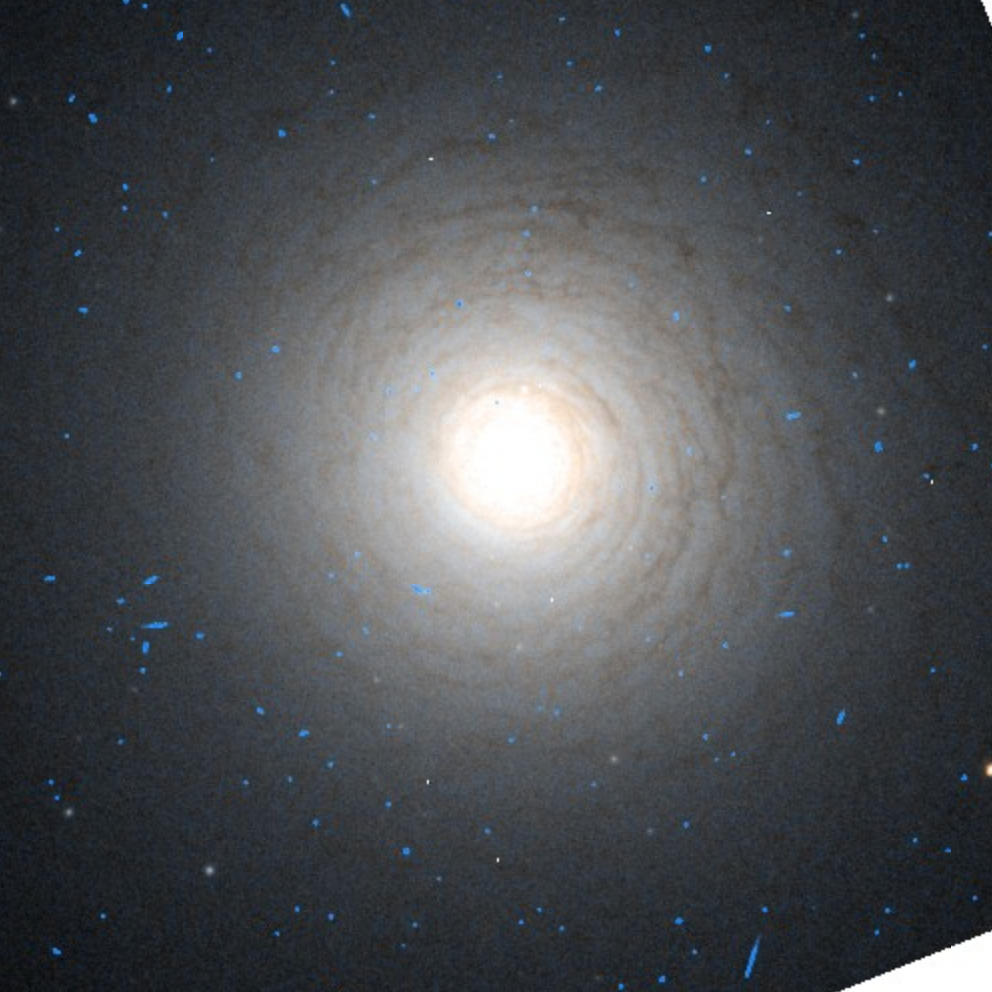} &
\includegraphics[trim=0.0cm 0cm 0.0cm 0cm, height=0.3\textwidth, angle=0]{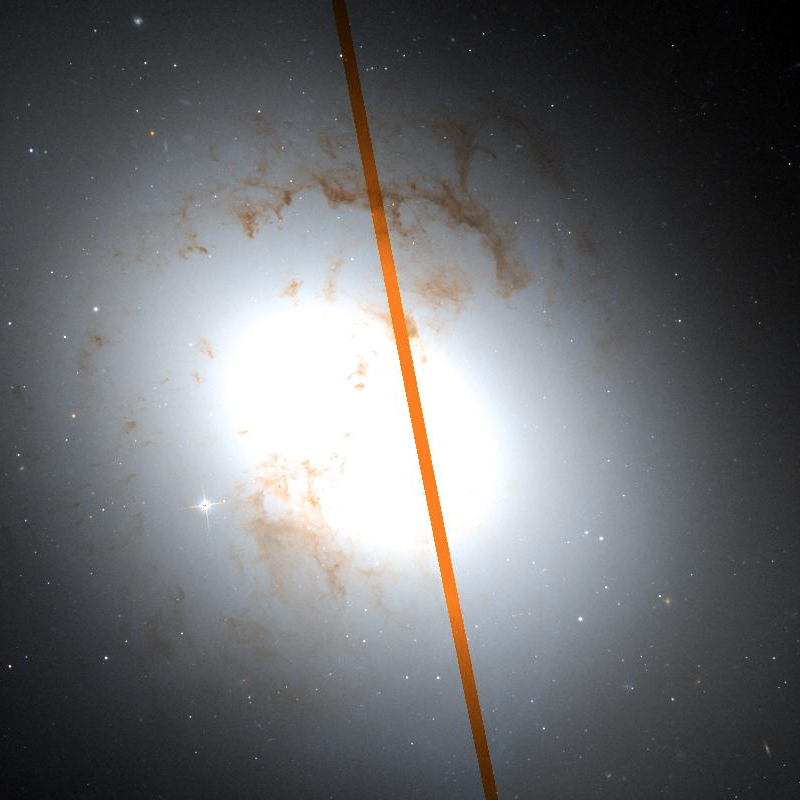} & 
\includegraphics[trim=0.0cm 0cm 0.0cm 0cm, height=0.3\textwidth, angle=0]{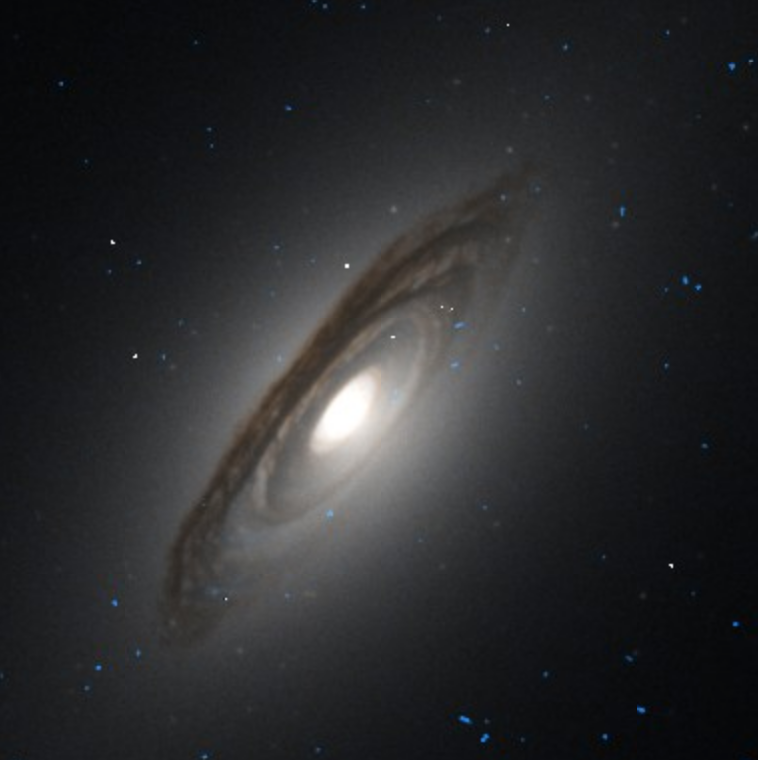} \\
\end{array} 
$
\end{center}
\caption{Examples of the four `dust bins' (dust code: N = strong No; n = weak no, only nuclear
  dust ring/disc; y = weak yes; Y = strong Yes) applied in Table~\ref{Table-dust}. 
Upper-left: (N)o visible dust. NGC~2778, inner $\sim$2.2$\times$2.2~kpc (HST
Obs.\ ID 6099).  Upper-middle: (n)uclear dust disc only. NGC~5845, inner
$\sim$4.4$\times$4.4~kpc with 0.9x0.9~kpc inset (Obs.\ ID 6099).  Upper-right:
(y)es dust but not much. NGC~5813, inner $\sim$2.9$\times$2.9~kpc (Obs.\ ID
5454).
Lower-left: (Y)es plenty of dust. NGC~524, inner $\sim$3.5$\times$3.5~kpc
(Obs.\ ID 5999).  Lower-middle: (Y)es widespread dust. NGC~1316,
$\sim$14.2$\times$14.2~kpc (Obs.\ ID 9409).  Lower-right: (Y)es plenty of
dust. NGC~6861, inner $\sim$2.6$\times$2.6~kpc (Obs.\ ID 5999).  Images are
central cutouts (not the whole galaxy) 
from F814W/F555W WFPC2/PC images available at the HLA, except for
NGC~1316, which is also from the HLA but is an F814W/F435W ACS/WFC image with
conjoined twin CCDs.  }
\label{Fig-dust}
\end{figure*}

\begin{table}
\centering
\caption{40 Lenticular (S0) and ellicular (ES,b) galaxies}\label{Table-data}
\begin{tabular}{llccc} 
\hline
Galaxy               &  Type & $\log(M_{\rm bh}/M_\odot)$ & $\log(M_{\rm *,sph}/M_\odot)$  &  $\log(M_{\rm *,gal}/M_\odot)$  \\
\hline 
\multicolumn{5}{c}{20 galaxies left of the S0/ES,b galaxy $M_{\rm bh}$-$M_{\rm *,gal}$ relation} \\
NGC 0404$^{\dagger}$         &  S0   &  5.74$\pm$0.10   &  8.03$\pm$0.50   & 9.19$\pm$0.16  \\
NGC 1332                     &  ES,b &  9.15$\pm$0.07   &  11.15$\pm$0.15  & 11.17$\pm$0.14 \\
NGC 1374                     &  S0   &  8.76$\pm$0.05   &  10.30$\pm$0.16  & 10.59$\pm$0.13 \\
NGC 2549                     &  S0   &  7.15$\pm$0.60   &  9.67$\pm$0.19   & 10.21$\pm$0.17 \\ 
NGC 2778                     &  S0   &  7.18$\pm$0.35   &  9.49$\pm$0.23   & 10.15$\pm$0.17 \\
NGC 2787$^{\dagger}$         &  S0   &  7.59$\pm$0.09   &  9.37$\pm$0.24   & 10.10$\pm$0.19 \\
NGC 3115                     &  ES,b &  8.94$\pm$0.25   &  10.87$\pm$0.14  & 10.95$\pm$0.13 \\
NGC 3245                     &  S0   &  8.30$\pm$0.12   &  10.12$\pm$0.17  & 10.70$\pm$0.15 \\
NGC 3998                     &  S0   &  8.33$\pm$0.43   &  10.12$\pm$0.26  & 10.61$\pm$0.15 \\
NGC 4026                     &  S0   &  8.26$\pm$0.12   &  10.19$\pm$0.22  & 10.44$\pm$0.17 \\
NGC 4339                     &  S0   &  7.62$\pm$0.33   &  9.73$\pm$0.21   & 10.22$\pm$0.14 \\
NGC 4342$^{\dagger}$         &  S0   &  8.65$\pm$0.18   &  10.04$\pm$0.15  & 10.36$\pm$0.15 \\
NGC 4350                     &  S0   &  8.87$\pm$0.41   &  10.39$\pm$0.25  & 10.66$\pm$0.13 \\
NGC 4434                     &  S0   &  7.85$\pm$0.17   &  9.95$\pm$0.20   & 10.22$\pm$0.13 \\
NGC 4564                     &  S0   &  7.77$\pm$0.06   &  10.08$\pm$0.16  & 10.35$\pm$0.14 \\
NGC 4578                     &  S0   &  7.28$\pm$0.35   &  9.79$\pm$0.15   & 10.24$\pm$0.13 \\
NGC 4742                     &  S0   &  7.13$\pm$0.18   &   9.78$\pm$0.16  &  10.07$\pm$0.14 \\
NGC 5845                     &  ES,b &  8.41$\pm$0.22   &  10.26$\pm$0.21  & 10.46$\pm$0.15 \\
NGC 6861                     &  ES,b &  9.30$\pm$0.08   &  11.07$\pm$0.19  & 11.15$\pm$0.18 \\
NGC 7457$^{\dagger}$         &  S0   &  6.96$\pm$0.30   &  9.34$\pm$0.17   & 10.12$\pm$0.15 \\
\multicolumn{5}{c}{20 galaxies right of the S0/ES,b galaxy $M_{\rm bh}$-$M_{\rm *,gal}$ relation} \\ 
NGC 0524$^{\ddagger}$ &  c-S0 &  9.00$\pm$0.10   &  10.88$\pm$0.15  & 11.38$\pm$0.13 \\
NGC 1023                     &  S0   &  7.62$\pm$0.05   &  10.33$\pm$0.16  & 10.89$\pm$0.14 \\
NGC 1194$^{\dagger\ddagger}$ &  m/S0 &  7.82$\pm$0.04   &  10.78$\pm$0.16  & 11.01$\pm$0.14 \\
NGC 1316$^{\dagger\ddagger}$ &  m/S0 &  8.16$\pm$0.29   &  11.05$\pm$0.35  & 11.69$\pm$0.31 \\
NGC 2974                     &  S0   &  8.23$\pm$0.07   &  10.48$\pm$0.22  & 10.98$\pm$0.16 \\
NGC 3384                     &  S0   &  7.23$\pm$0.05   &  10.14$\pm$0.16  & 10.67$\pm$0.14 \\
NGC 3489                     &  S0   &  6.76$\pm$0.07   &  9.53$\pm$0.20   & 10.30$\pm$0.14 \\
NGC 3665                     &  S0   &  8.76$\pm$0.10   &  11.14$\pm$0.16  & 11.39$\pm$0.14 \\
NGC 4371                     &  S0   &  6.84$\pm$0.12   &  9.99$\pm$0.30   & 10.70$\pm$0.22 \\
NGC 4429                     &  S0   &  8.18$\pm$0.08   &  10.60$\pm$0.20  & 11.04$\pm$0.13 \\
NGC 4459                     &  S0   &  7.82$\pm$0.10   &  10.56$\pm$0.21  & 10.77$\pm$0.15 \\
NGC 4526                     &  S0   &  8.65$\pm$0.04   &  10.79$\pm$0.26  & 11.13$\pm$0.15 \\
NGC 4594                     &  S0   &  8.81$\pm$0.03   &  11.04$\pm$0.25  & 11.26$\pm$0.13 \\
NGC 4596                     &  S0   &  7.90$\pm$0.20   &  10.28$\pm$0.20  & 10.86$\pm$0.13 \\
NGC 4762                     &  S0   &  7.24$\pm$0.14   &  9.74$\pm$0.15   & 10.83$\pm$0.13 \\
NGC 5018$^{\dagger\ddagger}$ &  m/S0 &  8.00$\pm$0.08   &  10.93$\pm$0.16  & 11.31$\pm$0.14 \\
NGC 5128$^{\dagger\ddagger}$ &  m/S0 &  7.65$\pm$0.12   &  10.71$\pm$0.25  & 11.14$\pm$0.12 \\
NGC 5252                     &  S0   &  9.03$\pm$0.40   &  10.97$\pm$0.27  & 11.50$\pm$0.15 \\
NGC 5813$^{\ddagger}$        &  c-S0 &  8.83$\pm$0.06   &  10.96$\pm$0.16  & 11.34$\pm$0.14 \\
NGC 7332                     &  S0   &  7.06$\pm$0.20   &  10.17$\pm$0.17  & 10.79$\pm$0.15 \\
\hline
\end{tabular}

Column~1: Galaxy name. 
$^{\dagger}$ Excluded from the Bayesian regression analyses in
\citet{Graham-sigma} for the reasons reiterated in Section~\ref{Sec_prev}. 
$^{\ddagger}$ Pre-recognised merger. 
Column~2: Galaxy Type. 
S0 = lenticular; c-S0 = merger-built core-S\'ersic lenticular; 
m/S0 = Previously recognised (wet merger)-built lenticular; 
ES,b = ellicular with possibly relic `red-nugget' bulge \citep{Graham:Sahu:22b}.  
Columns~3, 4, and 5: Black hole mass, spheroid stellar mass, and galaxy
stellar mass \citep[][and references therein]{Graham:Sahu:22a}. 
\end{table}

\section{Data set}\label{Sec_data}

\citet{Graham:Sahu:22a} describe a sample of 104 galaxies with directly
measured black hole masses and multicomponent decompositions of 
images taken at 3.6~$\mu$m
by the Spitzer Space Telescope and analysed by 
\citet{2016ApJS..222...10S}, 
\citet{2019ApJ...873...85D}, 
\citet{2019ApJ...876..155S}, and 
\citet{Graham:Sahu:22b}.  The galaxies' black hole masses were taken from the
literature, as indicated in the above papers.  As noted in 
\citet[][their Section~2.1]{2019ApJ...873...85D} and 
\citet[][their Table~4]{2019ApJ...876..155S}, the 
black hole masses are routinely adjusted to reflect their updated distances. The 
uncertainty on these black 
hole masses increased to incorporate the (random error) uncertainty in those
distances \citep[][their Table~1]{Graham:Sahu:22a}.  The distances were
obtained from a range of methods, thereby helping to erase the potential for 
an unknown systematic 
error in the distance --- introduced by individual distance indicator methods ---
that might produce a
bulk systematic bias in the masses of the black holes (and
galaxies).\footnote{In late 2022, Martin Bureau reminded the author that the
  community tend to overlook the two (random and systematic) uncertainties in 
the distance when reporting the uncertainty on the black hole mass. These should
be considered in addition to the typically reported random error on the black hole mass
measurement, which the author notes often does not include an allowance for systematic error
in its measurement from, for example, issues around spatial 
resolution or radially varying stellar populations due to galaxy components.}

Having noted the above, many of the ETG distances were based on their surface
brightness fluctuations \citep{2001ApJ...546..681T, 2002MNRAS.330..443B,
  2009ApJ...694..556B}, which did recently experience a systematic correction
of 1 per cent due to a reduced distance modulus for the Large Magellanic Cloud
\citep{2019Natur.567..200P}.  As such, their black hole masses shifted by 1
per cent.  As done by \citet{1996ApJ...463...26K} and
\citet{2022MNRAS.516.4066L}, for example, systematic errors can be separated
from random errors.  With the existence of (galaxy morphology)-dependent
scaling relations, coupled with the use of the same distance indicator
technique for galaxies of the same morphological type, there might be an
erroneous, bulk systematic shift to the values of $M_{\rm bh}$ for each galaxy
type.  This unknown systematic error can readily be attached as an additional
uncertainty on the intercept of the (galaxy morphology)-specific $M_{\rm
  bh}$-(stellar velocity dispersion, $\sigma$) relations if the same distance
indicator method was used for the galaxies that define each relation.
However, things are more convoluted when dealing with the $M_{\rm bh}$-$M_*$
relations given that systematic errors in the distance also impact $M_*$.
This leads to correlated errors in $M_{\rm bh}$ and $M_*$, coming from both
the random and systematic uncertainty in the distance.  Nonetheless, this
impact will be insignificant relative to the much larger factors of a few
difference in $M_{\rm bh}/M_*$, at fixed $M_*$, that are now observed between
the different morphological types.  Therefore, the issue of systematic errors
in the distance is left for a more thorough treatment elsewhere.

The data set also includes the galaxy morphological type, bulge stellar mass,
and galaxy stellar mass.  The sample consists of 73 
ETGs, of which 35 can be thought of as E galaxies\footnote{Technically, ten
  of these 35 are ellicular (ES,e) galaxies with intermediate-scale discs
  \citep{Graham:Sahu:22b}.}  and 38 can be thought of as S0
galaxies\footnote{Technically, four of these 38 are ellicular (ES,b) galaxies
  without large-scale discs \citep{Graham:Sahu:22b}.}, plus 31 LTGs, i.e.,
spiral galaxies, of which two are bulgeless (NGC~4395 and NGC~6926) and two
are reclassified here (and previously elsewhere) as S0 galaxies (NGC~2974 and
NGC~4594).  For convenience, masses and morphologies are provided in
Table~\ref{Table-data} for the S0 galaxies investigated here.

To be as inclusive as possible, \citet{2019ApJ...873...85D} considered
NGC~2974 and NGC~4594 (Sombrero) to be S galaxies.  However, as noted above,
they are now treated as S0 galaxies, thereby increasing the S0 count from 38
to 40.
NGC~4594 is classified in the literature as both an unbarred Sa galaxy and an
S0 galaxy.  With a $B/T$ flux ratio of 0.6, it had the highest spheroid mass
of all the 31 supposed S galaxies.\footnote{It is perhaps more akin to an ES
  galaxy, or more specifically, an ES,e galaxy \citep{Graham:Sahu:22b}, given
  that it appears to have an intermediate-scale disc rather than a large-scale
  disc, plus a low stellar density spheroid \citep{2019ApJ...873...85D}.}
The shell galaxy NGC~2974 \citep[aka NGC~2652:][]{2009AJ....138.1417T} is
often considered an S0 galaxy.  It has a weak inner bar and a nuclear,
gaseous, two-armed spiral structure within the inner 200 pc.  
It is thought to have accreted material and been built up by mergers with
smaller galaxies.  It has $0.7\times10^9$ M$_\odot$ of H{\footnotesize I}
\citep{1988ApJ...330..684K, 2008MNRAS.383.1343W} and dominates the isolated
NGC~2974 Group of five galaxies.

Hubble Space Telescope ({\it HST}) images, primarily `level 4' colour images
available from the Hubble Legacy Archive
(HLA)\footnote{\url{https://hla.stsci.edu}} of the 40 S0 galaxies, were
inspected for signs of dust.  The Mikulski Archive for Space Telescopes
(MAST)\footnote{\url{https://archive.stsci.edu}} was also used.  The S0
galaxies were assigned to one of four `dust bins', denoted as follows:
\begin{itemize}
\item N for (N)o visible sign of dust; 
\item n for (n)ot much other than a (n)uclear dust ring/disc and otherwise dust-free appearance; 
\item y for (y)es a little; and 
\item Y for a stronger (Y)es, with obvious signs of widespread dust beyond  the nucleus.
\end{itemize} 
Examples of the four dust bins are shown in Fig.~\ref{Fig-dust}. 
The outcome of the visual inspection is reported in Table~\ref{Table-dust},
along with a reference if dust was seen or if merger activity had previously
been reported. 

Some dusty S0 galaxies are well known to be young mergers (e.g., NGC~5128),
while others are old mergers which have since curtailed, if not ceased, their
star formation (e.g., NGC~1316).  It is expected that within the `down-sizing'
scenario, the lower-mass dusty S0 galaxies will, on average, have been built
most recently \citep{2020A&A...634A..95C, 2020ApJ...891L..23Z}.  What is of
relevance here is that they all experienced a major wet merger in the past,
evolving them to higher masses.

\subsection{Dust and the estimated $M_*/L_{3.6}$ ratios}\label{Sec_MonL}

This section explores how dust might contribute to an elevated derivation of
the stellar masses in the dusty S0 galaxies.  Indeed, dust can redden optical
colours, such as $B-V$, thereby increasing the ($B-V$ colour)-dependent
stellar mass-to-light ratios.  Based on the models of
\citet{2013MNRAS.430.2715I} and the \citet{2002Sci...295...82K} IMF,
\citep[][their Eq.~4]{Graham:Sahu:22a} present the following $M/L$ expression
for fluxes at 3.6 microns:
\begin{equation}
\log(M_*/L_{3.6}) = 1.034(B-V) - 1.067.
\label{Eq_MonL_IP13}
\end{equation}
From the $0.8 \lesssim B-V \lesssim 1.0$ colours of the S0 and ES galaxies
\citep[][their Fig.~1]{Graham:Sahu:22a}, none have unusually red
colours. However, perhaps some dusty S0 mergers have formed a younger
population in sufficient numbers that a more appropriate (dust-free) colour
would be more blue. If the dust-free colour ranged from $0.65 \lesssim B-V
\lesssim 0.9$, as seen for the LTGs, then the $M/L$ ratios would be $0.4
\lesssim M_*/L_{3.6} \lesssim 0.7$ rather than $0.6 \lesssim M_*/L_{3.6}
\lesssim 0.9$, with the latter based on the observed $0.8 \lesssim B-V
\lesssim 1.0$.  Dropping from a mid-point $M_*/L_{3.6}$ value of 0.75 to a
midpoint of $M_*/L_{3.6}=0.55$ would result in a $\sim$27 per cent reduction
in the mass-to-light ratio, resulting in a $\sim$0.135 dex decrease to the
stellar mass of the dusty, possibly intrinsically bluer, S0 galaxies.
However, the applicability of this adjustment is debatable given that the
$M/L$ formula was constructed using realistic dusty models for galaxies with a
range of morphologies.

Dust can also brighten the observed flux at 3.6~$\mu$m, which, although still
dominated by the stellar continuum, can contain the thermal glow of warm
dust. For this reason, \citet{2015ApJS..219....5Q} used a 25 per cent lower
$M_*/L_{3.6}$ ratio for LTGs than for ETGs. Therefore, one could argue for an
additional $\sim$0.125 dex reduction to the stellar masses of (some of) the
dusty S0 galaxies (if ongoing star formation is heating a substantial amount
of dust), giving a total 0.26~dex ($\approx$82 per cent) reduction when
combined with the reduction mentioned above based on dust reddening.  As we
shall see later, this is insufficient to explain the upcoming offset between
the dust-rich and dust-poor S0 galaxies.

\subsection{Exclusions}\label{Sec_Excl}

Readers not interested in the details of the sample can skip to
Section~\ref{Sec_Results}.  In what follows is a description of previously and
currently excluded galaxies by the author and the reasons for the exclusion.
Section~\ref{Sec_bulge} then provides a somewhat esoteric but relevant
discussion about the bulge masses of a few S and S0 galaxies in an arguably
undermining, albeit insufficiently so, manner not typically seen in black hole
scaling relation papers.

Identifying outlying data points is important for a couple of reasons.  Their
inclusion in regression analyses can yield a biased relation relative to that
defined by the bulk of the remaining population, especially if the outliers
are located towards the end of a distribution.  Their offset may be due to
measurement error, or it may signal that some additional process, such as
stripping, has modified the system.  As such, removing the outlying data is
appropriate and also serves as a means of flagging it for future
study. Suppose a relation is curved or bent when including more data beyond
the current sampling.  In that case, removing the apparent outlier(s) at the
extremity of what currently appears to be a linear (or log-linear)
distribution may enable the recovery of a more optimal approximation over that
fitted data range.

\subsubsection{Previous exclusions}\label{Sec_prev}

\citet{2013ARA&A..51..511K} used 17 and excluded 26 bulges from their sample
of 43 disc galaxies, declaring the bulk of the exclusions to be pseudobulges
which do not obey the near-linear $M_{\rm bh}$-$M_{\rm *,sph}$ relation.
Among what they thought were elliptical galaxies, they additionally excluded
classical bulges in known mergers, galaxies with depleted cores, and galaxies
with big black holes, basically systems which did not follow the near-linear
$M_{\rm bh}$-$M_{\rm *,sph}$ relation.

\begin{figure*}
\begin{center}
$
\begin{array}{cc} 
\includegraphics[trim=0.0cm 0cm 0.0cm 0cm, width=1.0\columnwidth, angle=0]{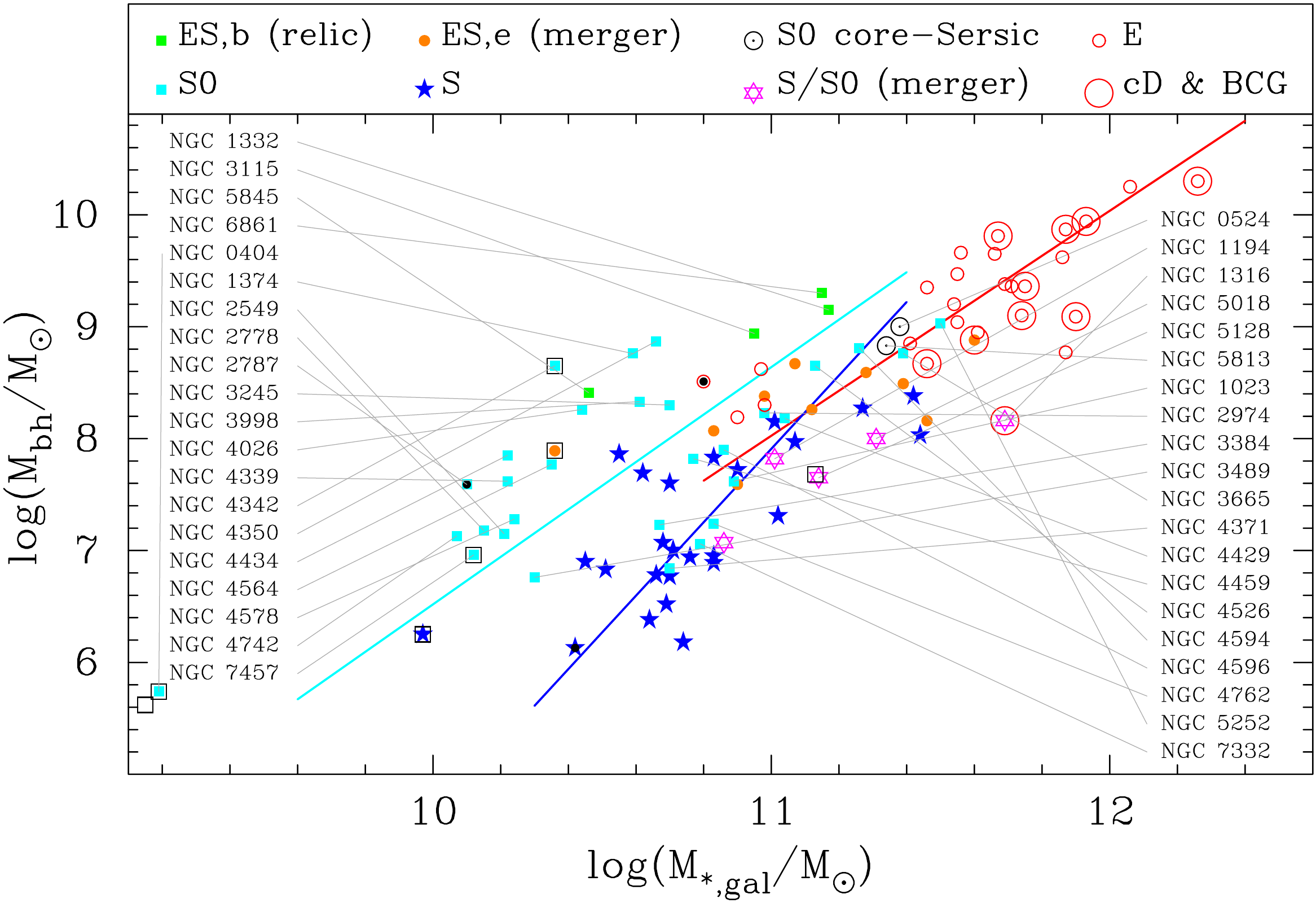} & 
\includegraphics[trim=0.0cm 0cm 0.0cm 0cm, width=1.0\columnwidth, angle=0]{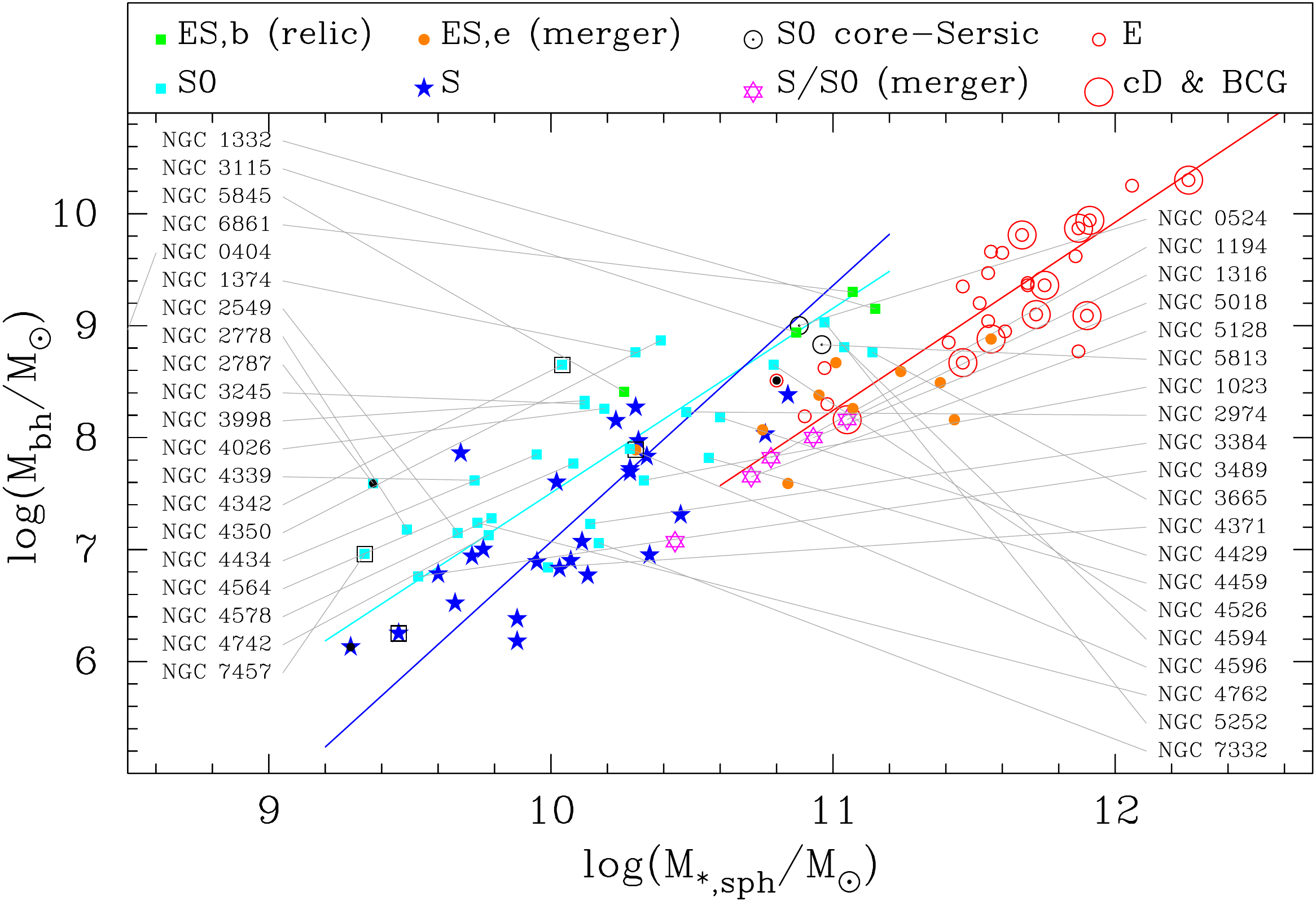} \\
\end{array} 
$
\end{center}
\caption{Left: Modification of Fig.~2 from \citet{Graham-sigma}.
Black hole mass versus galaxy stellar mass.
Five points enclosed in a black square (three S0, the S galaxy
Circinus S, and the ES.e galaxy NGC~3377 mentioned in 
Section~\ref{Sec_prev}), plus three points over-plotted with a black
circle (the S0 galaxy NGC~2787, the E galaxy NGC~4291, and the S galaxy NGC~4945),
were excluded from the Bayesian regression analyses performed in
\citet{Graham-sigma}, as were the bulgeless galaxies NGC~4395 and NGC~6926,
denoted by the empty black squares in the left panel. 
The regressions shown here were for the E, ES,e and two core-S\'ersic galaxies (red
line),  the S0 and ES,b galaxies (cyan line), 
and the S galaxies (blue line).  
Right: Similar, but using the spheroid stellar
  mass rather than the galaxy stellar mass.
The regressions sample the same systems 
in both panels.  For plotting purposes only, NGC~2974 (aka NGC~2652) 
and NGC~4594 (M104) have been reclassified here from S to S0. 
}
\label{Fig-M-S0-gal}
\end{figure*}

The current work instead builds on the (galaxy morphology type)-dependent
$M_{\rm bh}$-$M_{\rm *,sph}$ and $M_{\rm bh}$-$M_{\rm *,gal}$ relations most
recently presented in \citet{Graham-sigma}, where some galaxies were also
excluded from the Bayesian regression analyses performed there.\footnote{The
  Bayesian analyses is described in the Appendix of
  \citet{2019ApJ...873...85D}.}  To maintain sample consistency in
\citet{Graham-sigma}, if a galaxy was flagged as an outlier in one diagram,
such as the $M_{\rm *,sph}$-$\sigma$ diagram, it was excluded from the
regression in all diagrams.  These excluded galaxies are shown in
Fig.~\ref{Fig-M-S0-gal}, except for the two bulgeless galaxies (NGC~4395 and
NGC~6926) and one galaxy with no black hole mass (NGC~5055)\footnote{The
  `black hole' mass is actually the entire mass interior to 300~pc
  \citep{2004A&A...420..147B}.  As such, it is not the black hole mass and is
  not included here.}, and they are discussed below.

Previously, five of the initial 104 galaxies (NGC~1194, NGC~1316, NGC~2960,
NGC~5018, and NGC~5128)\footnote{In passing, it is remarked that NGC~1947 (not
  in sample) may represent an intermediary between NGC~5128 and NGC~1316,
  while IC~1623 (not in sample) may represent an earlier stage of 
  NGC~5128.}  were {\it a priori} excluded due to their well-recognised wet
merger status.  Although the community, and author, had tended to discard such
galaxies as (unrelaxed) deviant systems that should be shunned from the
scaling diagrams, these four S0 galaxies plus one S galaxy (NGC~2960) have
yielded vital insight into understanding the (evolution of the) S0 galaxy
population.  These five galaxies are marked with pink hexagons in
Fig.~\ref{Fig-M-S0-gal}.  An additional disturbed S galaxy
\citep[Circinus:][]{1998MNRAS.300.1119E} --- marked in Fig.~\ref{Fig-M-S0-gal}
with a black square around a blue star --- was also excluded from the past
regression analyses for the same reason.

Three galaxies stood out as outliers from the morphology-dependent $M_{\rm
  *,sph}$--$\sigma$ relations \citep{Graham-sigma}. The outliers included one
E galaxy (NGC~4291), one dusty S0 galaxy (NGC~2787), and one S galaxy (NGC~4945),
and may represent measurement error.  As discussed in Appendix~B of
\citet{Graham-sigma}, NGC~4291 may be an S0 galaxy rather than an E galaxy and
therefore have an incorrectly high spheroid mass assigned to it.  NGC~4945 may
also have an erroneous spheroid mass (and spheroid S\'ersic index), while only
NGC~2787 remains something of a mystery.  These three galaxies have been
over-plotted with a black dot in Fig.~\ref{Fig-M-S0-gal}.

This leaves four more past exclusions.\footnote{It is five, rather than four,
  past exclusions when counting NGC~5055, which is excluded from the sample of
  104 because it is without a measured black hole mass.}  NGC~404 (S0) is
located at the low-mass end of the diagram, and as such, it could have
excessive weight in the regression analyses, torquing the fitted line towards
it.  It may, however, be in the correct location.  The bulgeless galaxy
NGC~4395 resides close to it in the $M_{\rm bh}$-$M_{\rm *,gal}$ diagram
(Fig.~\ref{Fig-M-S0-gal}).  Second, NGC~4342 is a stripped S0 galaxy, having
had its stellar mass reduced by an unknown amount.  The ES,e ellicular galaxy
NGC~3377 \citep{2015AdSpR..55.2372N}, also known as a discy fast-rotating
elliptical galaxy, was excluded due to its biasing nature in the $M_{\rm
  bh}$-$M_{\rm *,gal}$ diagram \citep[][his
  Fig.~2]{Graham-sigma}.\footnote{This bias would be reduced or disappear if
  NGC~3377 was an S0 galaxy.}  Finally, NGC~7457 (S0) has an unusual kinematic
structure, including cylindrical rotation about its major axis, signalling
that it too experienced a past merger \citep{2002ApJ...577..668S,
  2003ApJ...597..893N, 2019MNRAS.488.1012M}.

In Figure~\ref{Fig-M-S0-gal}, the $M_{\rm bh}$-$M_{\rm *,gal}$ and $M_{\rm
  bh}$-$M_{\rm *,sph}$ relations from \citet{Graham-sigma}, based on these
exclusions, are shown for the E, S0 and S galaxies. 

\subsubsection{Current exclusions}

Given the inclusion here of the above five mergers 
(NGC~1194, NGC~1316, NGC~2960, NGC~5018, and
NGC~5128) and the current focus on S0 (including ES,b) galaxies, 
only four S0 galaxies from the above list are 
excluded: 
NGC~404 (low-mass extremity); NGC~2787 (currently unexplained outlier); NGC~4342 (stripped); and
NGC~7457 (merger with unusual kinematics).  They are marked in all
relevant diagrams.  

In addition, based on the appearance of dust, as reported in
Table~\ref{Table-dust}, one dusty (NGC~3489) and one relatively dust-free S0
galaxy (NGC~1332) are flagged and excluded from the regression analyses.
These two systems are not thought to be in error but reflect that the merging
process is not a perfectly regimented sequence.  That is, a small fraction of
real outliers exist.  For example, the S0 galaxies NGC~1023, NGC~4762, and
NGC~7332 are not dusty but are, or may be, post-mergers (see the comments in
Table~\ref{Table-dust}).
These and a few other (included) systems are discussed further in
Section~\ref{Sec_noncon}.  While NGC~4945 --- the only previously excluded
spiral galaxy --- does not have an undue influence on the mass-mass diagrams
presented here and is therefore included, the spiral galaxy NGC~1300 was
flagged as an outlier in \citet{Graham:Sahu:22a} and has been excluded here
(see Fig.~\ref{Fig-6panel}) in deriving slightly revised S galaxy scaling
relations.

These galaxies are partly noted because, like the stripped galaxy NGC~4342,
they may offer further insight into the processes which mould galaxies.  They
may, however, turn out to have measurement errors or instead reflect true
scatter in the evolutionary sequence.

\subsubsection{Unknown exclusions and selection biases}

Disc-dominated galaxies with small
bulges, including low surface brightness disc galaxies with large disc
scalelengths \citep[e.g.][]{1982PASJ...34..423H, 2001ApJ...556..177G}, 
may be under-represented.  Indeed, several LTGs were excluded
from the parent sample established in \cite{Graham:Sahu:22a} because they 
needed to be analysed using Hubble Space Telescope data rather than Spitzer Space
Telescope data to resolve their smaller bulges.  However, their
inclusion by \citet{2018ApJ...869..113D} reveals a consistent trend in the
$M_{\rm bh}$-$M_{\rm *,gal}$ and 
$M_{\rm bh}$-$M_{\rm *,gal}$ diagram.  Nonetheless, one should be mindful of
potential selection bias
\citep{1976Natur.263..573D} in the $M_{\rm bh}$-$M_{\rm *,gal}$ diagram.

The previous concern of a sample selection bias between galaxies with directly
measured black hole masses and those without \citep{2016MNRAS.460.3119S} arose
from the use of an inconsistent stellar mass-to-light ratio between the
samples imaged in different bands.  In the $M_{\rm *,gal}$--$\sigma$ diagram,
ETGs with directly measured black hole masses do not appear offset from those
without a directly measured black hole mass \citep{SahuGrahamHon22}.

\subsection{What is a bulge?}\label{Sec_bulge}

Quantifying a bulge once meant performing a two-component bulge$+$disc
decomposition. However, closer inspection of nearby galaxies has revealed the
presence of many components that, when accounted for, can whittle away at
what was once considered the bulge.  This subsection is primarily included to
raise awareness of this issue, which previously received some impetus from the
three-component bulge$+$bar$+$disc fits in \citet{2005MNRAS.362.1319L}.
However, multicomponent fits with $R^{1/n}$ bulges were pioneered by
\citet{1997AJ....114.1413P} and \citet{2001A&A...367..405P}.

The bulge masses used here come from multicomponent decompositions in which
additional components, such as bars, ansae, rings, nuclear star clusters,
nuclear discs, nuclear bars, and more, were identified in images and sometimes
spectra.  In many instances, the components were already known in the
literature.  There is usually little ambiguity over the bulge, but sometimes,
there is scope for question. In pushing the frontiers of what is feasible, a
couple of examples are discussed below.

While NGC~2787 ($\log\,\sigma [{\rm km}\,{\rm s}^{-1}]=2.28$) has a light
profile --- shown in \citet[][]{Graham:Sahu:22b} --- which is similar to that
of NGC~4371 ($\log\,\sigma [{\rm km}\,{\rm s}^{-1}]=2.11$) --- shown in
\citet{2019ApJ...876..155S} --- the stellar mass of NGC~2787's spheroid and
galaxy is $\sim$4 times smaller.  Curiously, \citet{2015A&A...584A..90G}
reveal how the bulge component of NGC~4371 might be smaller and, indeed, how
it could be whittled away to nothing.  NGC~4371, like NGC~4429, is a
'figure-of-eight' galaxy, with its X-shaped `pseudobulge' structure lining up
with its partial ring (likely the ansae of a now dispersed bar) to form a
'figure-of-eight' pattern.
While \citet{2019ApJ...876..155S} fit a S\'ersic bulge plus a (low-$n$
S\'ersic) barlens component for the pseudobulge in NGC~4371,
\citet{2015A&A...584A..90G} fit a point-source, a nucleus (<0$\arcsec$.8),
plus an ``inner disc'' (likely dominating from 1 to 20$\arcsec$) for the
pseudobulge component. \citet{2015A&A...584A..90G} noted that their
pseudobulge component additionally encompassed unmodelled light from both a
3$\arcsec$.2 disc-like feature\footnote{Could this be the inner part of the
  bar?}  \citep[light which contributed to the bulge model
  of][]{2019ApJ...876..155S} and a 10$\arcsec$ ring \citep[which contributed
  to the barlens model of][]{2019ApJ...876..155S}.  Although this does not
impact the $M_{\rm bh}$-$M_{\rm *,gal}$ diagram explored here, it raises the
question, What is a bulge?  This can influence the $M_{\rm bh}$-$M_{\rm
  *,sph}$ diagram if nearby galaxies or galaxies with new, better spatial
resolution are decomposed {\it ad infinitum}.  It also raises the question as
to which component(s) dominate the stellar velocity dispersion, $\sigma$, used
in $M_{\rm bh}$-$\sigma$ diagrams.

The SB(s)c galaxy NGC~3079, with $M_{\rm bh}=2.4\times10^6$ M$_\odot$, is
another potentially interesting example in which some of its bulge might
be from a
`pseudobulge'.  The galaxy has $\log(M_{\rm *,gal}/M_\odot)=10.64\pm0.20$ dex,
and is reported as having $\log(M_{\rm *,sph}/M_\odot)=9.88\pm0.29$ dex with
$R_{\rm e,sph,equiv}=0.34\pm0.04$ kpc \citep{2019ApJ...873...85D,
  2018ApJ...869..113D, Graham:Sahu:22a}.  
This compares closely with the Milky Way, with morphological type SB(rs)bc and
a prominent (peanut shell)-shaped `pseudobulge' structure.  The Milky Way has
$M_{\rm bh}=4.0\times10^6$ M$_\odot$ \citep{2016ApJ...830...17B}, $\log(M_{\rm
  *,gal}/M_\odot)=10.78\pm0.10$ dex and $\log(M_{\rm
  *,sph}/M_\odot)=9.96\pm0.05$ dex \citep{2015ApJ...806...96L}.  The Milky
Way's expected half-light bulge radius is $\sim$0.5~kpc, based on the typical S galaxy
$R_{\rm e,sph,equiv}/h_{\rm disc}$ ratio of 0.2 \citep{1996ApJ...457L..73C,
  2008MNRAS.388.1708G} and the Milky Way's measured $h_{\rm disc}\approx 2.5$~kpc
\citep{2017MNRAS.471.3988C}.  This is smaller than the measured value $R_{\rm
  e,sph,equiv}=0.8$ kpc (5.85 degrees) from \citet{2007ApJ...655...77G},
perhaps biased by the pseudobulge.  Both galaxies (NGC~3079 and the Milky Way)
feature bipolar gas bubbles
\citep{1988ApJ...326..574D, 1994ApJ...433...48V, 2000ApJ...540..224S,
  2001ApJ...555..338C, 2003ApJ...582..246B}.  It is known that satellite
galaxies rain down on the Milky Way \citep{2021Galax...9...66P}, and this may
also be likely in NGC~3079, given that it resides in the N3078/N3079 Group with
another 18 less massive members.

There is also the curious case of the giant irregular galaxy NGC~6926, which,
at 86~Mpc, is the most distant S galaxy in the present sample.\footnote{Aside
  from the S galaxy merger NGC~2960 at 73~Mpc, and UGC~3789 at 51~Mpc, all of
  the sample's S galaxies are located within 40~Mpc.}  It is thought to be
disturbed by its dwarf ETG neighbour NGC~6929 and has a nuclear, molecular
disc thought to be edge-on and thus some 18 to 32 degrees misaligned with the
inclination of the main stellar disc \citep{2005PASJ...57..587S,
  2019ApJ...873...85D}.  Despite its large stellar mass, this S galaxy is
bulgeless.\footnote{NGC~6926 was modelled as having a pseudobulge rather than
  a classical bulge \citep{2019ApJ...873...85D}.}
NGC~4699 is another galaxy with an unexpectedly low bulge-to-total stellar
mass ratio, of just 0.10, for a massive spiral galaxy \citep{Graham:Sahu:22b}.  

\begin{figure*}
\begin{center}
\includegraphics[trim=0.0cm 0cm 0.0cm 0cm, width=1.0\textwidth, angle=0]{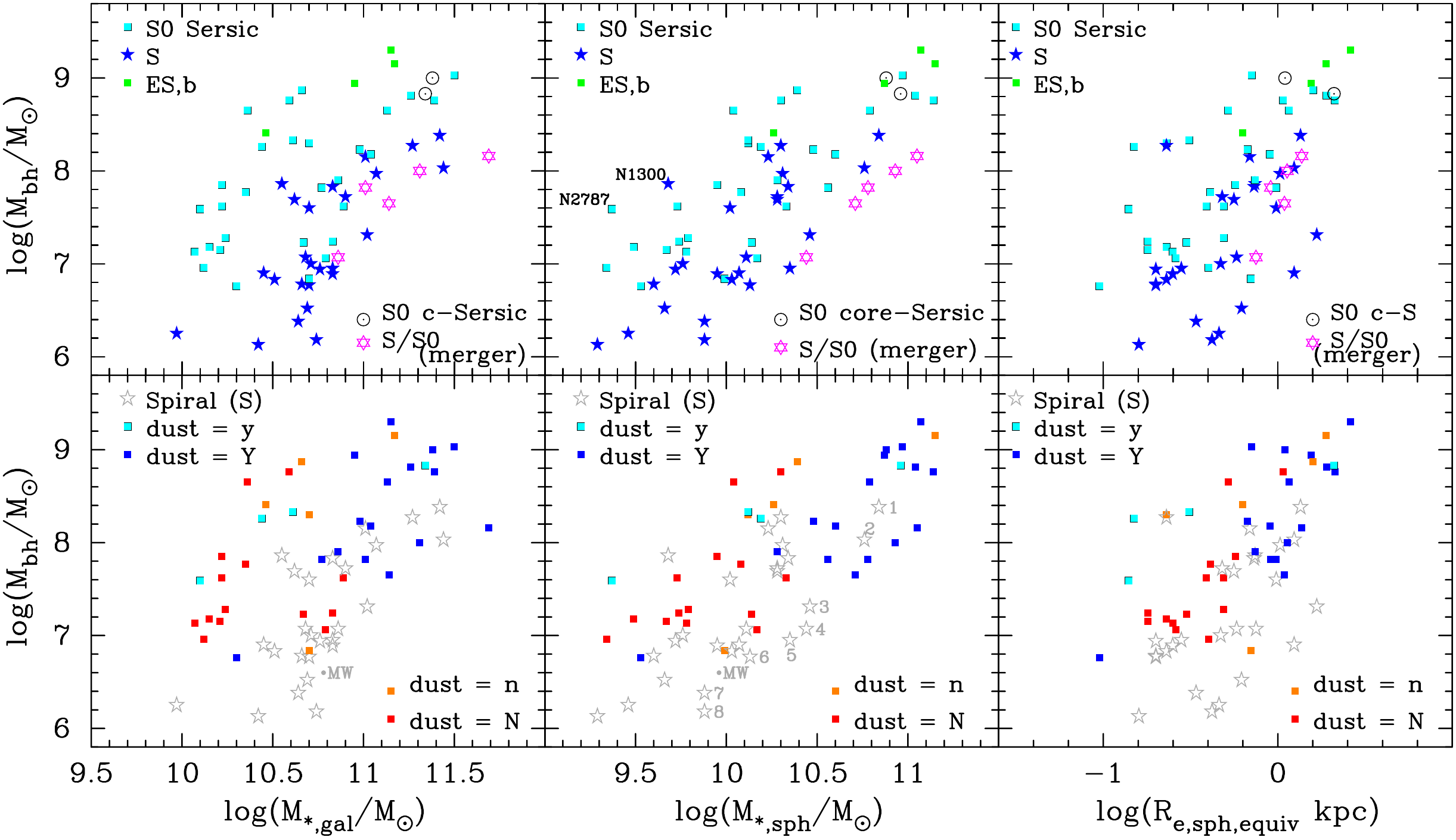}
\caption{Adapted from Fig.~2 in \citet{Graham-sigma}. 
Distribution of disc galaxies (S and S0, including core-S\'ersic
  S0, S0 mergers, and ES,b) in the 
$M_{\rm bh}$-$M_{\rm *,gal}$ (left), 
 $M_{\rm bh}$-$M_{\rm *,sph}$ (middle), and 
 $M_{\rm bh}$-$R_{\rm e,sph}$ (right) diagrams.
Dust code: (Y)es plenty, (y)es but not much, (N)othing or just a small (n)uclear dust
 disc/ring. 
Due to their location on the high $M_{\rm *,sph}/M_{\rm bh}$ envelope of the
distribution, several spiral galaxies are labelled in the lower-middle panel, they are: 
1 = NGC~1097; 2 = NGC~1398; 3 = NGC~4501;
4 = NGC~2960 (S/merger); 5 = NGC~2273; 
6 = NGC~1320; 7 = NGC~3079; 8 = NGC~4826 (aka the Black Eye
Galaxy); and MW = the Milky Way galaxy, with a bulge radius 
of 0.5--0.8~kpc (not plotted).  
}
\label{Fig-6panel}
\end{center}
\end{figure*}

While systems with ambiguous bulges are thought to represent a minority of the
sample, Fig.~\ref{Fig-6panel} presents not only the $M_{\rm bh}$--$M_{\rm
  *,sph}$ diagram (and $M_{\rm bh}$--$R_{\rm e,sph}$ diagram) but also the
$M_{\rm bh}$--$M_{\rm *,gal}$ diagram for the S0 galaxies, along with the S
galaxies.  The `dust bin' of each S0 galaxy is indicated there.  Given the
focus on the S0 galaxies, the S galaxies are greyed out but retained for
reference.  The `dust bin' to which each S0 galaxy was assigned can also be
seen in Table~\ref{Table-dust}.  It is apparent (from the $M_{\rm
  bh}$--$M_{\rm *,gal}$ diagram) that the division among the S0 galaxies is
not due to the decomposition process.

\begin{table*}
\centering
\caption{Dust and comments on the S0 and ES,b galaxies}\label{Table-dust}
\begin{tabular}{lll} 
\hline
Galaxy                       &  Dust   & Comments \\
\hline 
\multicolumn{3}{c}{19 largely non-dusty (excluding small nuclear dust rings/discs) S0 galaxies} \\
NGC 1023                     &  N, n   & Disturbed H{\footnotesize I} disc 
and possible minor merger \citep{2012MNRAS.423.2957B} \\ 
NGC 1332                     &  n, ... & Thin nuclear dust disc at $r\lesssim$240~pc \citep{2016ApJ...823...51B}. \\
NGC 1374                     &  N, ... &  \\
NGC 2549                     &  N, n   &  \\ 
NGC 2778                     &  N, n   &  \\ 
NGC 3245                     &  n/y, n  & Nuclear dust rings at
$r\lesssim300$~pc plus snakey dust lane \citep{2008AJ....135..747G, 2010MNRAS.402.2462C}. \\ 
NGC 3384                     &  N, n   &  Possibly starting to interact with external H{\footnotesize I} gas \citep{2003ApJ...591..185S}. \\
NGC 4339                     &  N, n   &  \\
NGC 4342$^{\dagger}$         &  N, n   &  Stripped and within a quenching X-ray halo \citep{2014MNRAS.439.2420B}. \\
NGC 4350                     &  n, n   & Thin nuclear dust disc/ring with radial extent $r\sim$330~pc \citep{1997AJ....113..950F}. \\
NGC 4371                     &  n, n   & Nuclear dust disc/ring at $\sim$50~pc
and very thin star-forming ring at $\sim$0.83~kpc \citep{2010MNRAS.402.2462C}. \\ 
NGC 4434                     &  N, n   &  \\
NGC 4564                     &  N, n   &  \\
NGC 4578                     &  N, n   &  \\
NGC 4742                     &  N, ... &  \\ 
NGC 4762                     &  N, n   & Thick asymmetric gas disc plus warped stellar disc \citep{1994AnA...286L...5W}. \\ 
NGC 5845                     &  n, d   & Nuclear dust disc/ring at
$r\sim$120~pc \citep{2006ApJ...640..143S, 2012ApJ...749L..10J}. \\
NGC 7332                     &  N, n   & Small counter-rotating gas disc \citep{1994AJ....107..160F, 1996AnA...307..391P, 1999AJ....117.2725S, 2004MNRAS.350...35F}. \\
NGC 7457$^{\dagger}$         &  N/n, n   & Faint dust ring at $r\sim3.2$~kpc. Possible unequal-mass merger \citep{2011ApJ...738..113H}. \\
\multicolumn{3}{c}{21 dusty S0 galaxies} \\ 
NGC 0404$^{\dagger}$         &  Y, ... & Dust lane, extensive gas disc/doughnut, accretion 0.5~Gyr ago, central star-forming UV ring \citep{2004AJ....128...89D, 2010AnA...513A..54B}. \\ 
NGC 0524$^{\ddagger}$        &  Y, n   &  Extensive dust rings. Merger,
core-S\'ersic light profile \citep{2009MNRAS.399.1839K}. \\ 
NGC 1194$^{\dagger\ddagger}$ &  Y, ... &  Dust ring. Merger and possibly
interacting with PGC 1127439. \citep{2000ApJS..128..479S}. \\
NGC 1316$^{\dagger\ddagger}$ &  Y, ... & Dusty 3 Gyr old merger \citep{2001MNRAS.328..237G, 2019AnA...628A.122S}. \\
NGC 2787$^{\dagger}$         &  y, ... & Misaligned dust rings/disc $r\lesssim$160~pc plus an outer H{\footnotesize I} ring \citep{2004AJ....127.2641S, 2010MNRAS.402.2462C}. \\ 
NGC 2974                     &  Y, n   & Dust clearly visible \citep{2005MNRAS.357.1113K}. Star formation 1 Gyr ago \citep{2007MNRAS.376.1021J}. \\
NGC 3115                     &  Y, ... & Edge-on disc. Hot gas core \citep{2014ApJ...780....9W, 2021MNRAS.504.2146B}. \\ 
NGC 3489                     &  Y, d   & Dust lane \citep{2003ApJ...584..260W}. \\  
NGC 3665                     &  Y, d   & High gas content, inner dust disc  \citep{2013MNRAS.432.1796A}. \\ 
NGC 3998                     &  y, n   & Faint dust lanes and nuclear star-formation, warped large-scale H{\footnotesize I} disc \citep{1985AnA...142....1K, 1991AJ....101..102W, 2016AnA...592A..94F}. \\
NGC 4026                     &  y, n   & Faint (misaligned) dust lanes; accreting a little H{\footnotesize I} \citep{1988AnA...199...41V}. \\
NGC 4429                     &  Y, d   & Accreted inner disc \citep{2013MNRAS.432.1796A, 2015ApJS..216....9C}.\\
NGC 4459                     &  Y, d   & Prominent inner dust disc \citep{2009AJ....137.3053Y}.  Unbarred, double nuclear ring \citep{2008ApJ...676..317Y, 2010MNRAS.402.2462C}. \\ 
NGC 4526                     &  Y, d   & Prominent inner dust-disc (aka Black Eye Galaxy, NGC~4560) \citep{2000AJ....120..123T}. \\ 
NGC 4594                     &  Y, ... & Sombrero galaxy, aka M104 \citep{1996AnA...312..777E}. \\
NGC 4596                     &  Y, n   & Inner dust disc \citep{1990AJ....100..377K, 1999MNRAS.306..926G}. \\ 
NGC 5018$^{\dagger\ddagger}$ &  Y, ... & Dust lanes, merger, shells, gas-bridge to NGC~5022 \citep{1996AnA...314..357H, 2009AJ....138.1417T, 2018ApJ...864..149S}. \\ 
NGC 5128$^{\dagger\ddagger}$ &  Y, ... & Dusty gas-rich merger \citep{1983ApJ...272L...5M, 1998AnARv...8..237I}. \\ 
NGC 5252                     &  Y, ... & Merger, misaligned dust lanes \citep{1998ApJ...505..159M, 2015AJ....149..155K, 2015ApJ...814....8K}. \\ 
NGC 5813$^{\ddagger}$        &  y, n   & Widespread dust, old merger, core-S\'ersic galaxy, misaligned kinematic axis, X-ray halo  \citep{2000AJ....120..123T, 2015MNRAS.452....2K}. \\ 
NGC 6861                     &  Y, ... & Dusty gas disc; interacting with
NGC~6868, and located near IC~4943 \citep{2010ApJ...711.1316M}. \\ 
\hline
\end{tabular}

Column~1: Same as Table~\ref{Table-data}. 
Column~2: Is there dust visible in the HST optical image: 
(Y)es plenty; (y)es but not a lot; (n)ot really, just a (n)uclear dust
disc/ring; and (N)o, none. 
The second entry, after the comma, is the ``Dust feature'' classification from \citet{2011MNRAS.414.2923K}: 
n = no dust; d = dusty disc.
Column~3: Comments and references showing the dust and/or noting accretion/merger
activity. 
\end{table*}

\section{Results}\label{Sec_Results}

\subsection{Splitting the lenticular galaxies}
\label{Sec_S0_split}

Compared to S galaxies, S0 galaxies exhibit a considerable amount of scatter
in the $M_{\rm bh}$-$M_{\rm *,gal}$, and $M_{\rm bh}$-$M_{\rm *,sph}$ diagrams
(Fig.~\ref{Fig-M-S0-gal}).  An exploration of this scatter was the motivation
for this paper.

As noted in Section~\ref{Sec_data}, a visual inspection was performed on the
sample of S0 galaxies.  Getting one's hands dirty and a little more intimate
with the galaxies proved highly informative.  Doing so, coupled with recourse
to the literature, a general division emerged between the S0 galaxies living
on the right-hand and left-hand side of the S0 galaxy distribution in the
$M_{\rm bh}$-$M_{\rm *,gal}$ and $M_{\rm bh}$-$M_{\rm *,sph}$ diagrams,
evident in Fig.~\ref{Fig-6panel}.  By and large, those S0 galaxies on the
left-hand side do not display much evidence for dust, a tracer indicative of
an environment suitable for hosting cold gas.  In contrast, most of those on
the right-hand side abound in dust and tend to be previously-recognised wet
mergers (Table~\ref{Table-dust}).  Furthermore, after completing this work, it
was discovered that 
the notion of two S0 galaxy subtypes was realised by \citet{1990ApJ...348...57V} via an 
exploration of the galaxy luminosity functions and ellipticity histograms,
although the S0 population remained somewhat enigmatic \citep{2009ApJ...702.1502V}.

Fig.~\ref{Fig-dust} shows an illustrative `facebook' for some S0 galaxies in
the different dust bins.  Additional examples of dusty S0 galaxies can be seen
in Fig.~B1 of \citet{2013MNRAS.433.2812K} and
Dustpedia\footnote{\url{www.dustpedia.com}} \citep{2017PASP..129d4102D,
  2018A&A...609A..37C}.  Upon commencing this project, only four of the
sample's S0 galaxies had been flagged as recognised mergers.  Appearing as low
$M_{\rm bh}/M_{\rm *,sph}$ outliers from past near-linear $M_{\rm bh}$-$M_{\rm
  *,sph}$ relations, they were previously identified as mergers and dismissed
as presumably unrelaxed systems.  However, they need not be dismissed, nor are
they rare; they are simply the lower $M_{\rm bh}/M_{\rm *,sph}$ ratio members
of a more extensive distribution of dusty S0 galaxies built by substantial wet
mergers.

\subsubsection{New S0 galaxy $M_{\rm bh}$-$M_*$ scaling relations}

Given the trends in Fig.~\ref{Fig-6panel}, the previous $M_{\rm bh}$-$M_{\rm
  *,sph}$ relations for S0 and S galaxies from \citet{Graham-sigma} are
revisited in Fig.~\ref{Fig-MMsph}, where separate relations for the dust-poor
and dust-rich S0 galaxies are reported, along with a slightly revised relation
for the S galaxies due to the changes mentioned in Section~\ref{Sec_data}.
What can be seen in Fig.~\ref{Fig-MMsph} is tantamount to a minor breakthrough
in understanding the connection between galaxies.  It had been a mystery why
the S galaxies followed a steeper relation than the ETGs
\citep[][Fig.~\ref{Fig-M-S0-gal}]{2016ApJ...817...21S, 2019ApJ...873...85D}
and how it arose that the S galaxies defined a steep $M_{\rm bh}$-$M_{\rm
  *,sph}$ relation {\it between} the S0 and E galaxies
\citep{2019ApJ...876..155S}.  Fig.~\ref{Fig-6panel} and \ref{Fig-MMsph}
reveal, for the first time, that the dust-poor and dust-rich S0 galaxies also
follow steep $M_{\rm bh}$-$M_{\rm *,sph}$ relations, which sandwich the S
galaxies.  The dust-poor S0 galaxies define the relation
\begin{equation}
\log(M_{\rm bh}/M_\odot)=
(2.39\pm0.81)[\log(M_{\rm *,sph}/M_\odot)-10.00]+(7.67\pm0.26), 
\end{equation} 
at least for $7 < \log(M_{\rm bh}/M_\odot) < 9$ dex.  An expression with a
smaller uncertainty at a different intercept is provided in the caption to
Fig.~\ref{Fig-MMsph}.  At a given black hole mass, the dust-rich S0 galaxies
have a spheroid mass that is, on average, $\sim$0.54 dex greater than that of
the dust-poor S0 galaxies in the sample.  The Bayesian regression hints at a
steeper slope for the dust-rich S0 galaxies; however, the uncertainty is
sufficiently large that this remains inconclusive.  Nonetheless, the intercept
of the dust-rich S0 galaxy $M_{\rm bh}$-$M_{\rm *,sph}$ relation (for 15
galaxies excluding NGC~3489) appears notably different in
Fig.~\ref{Fig-MMsph}, with
\begin{equation}
\log(M_{\rm bh}/M_\odot)= 
(3.69\pm1.51)[\log(M_{\rm *,sph}/M_\odot)-10.70]+(8.42\pm0.15), 
\label{Eq-xtc}
\end{equation}
noting that a different normalisation point, at $\log(M_{\rm
  *,sph}/M_\odot)=10.70$ dex, has been used in Eq.~\ref{Eq-xtc}.

Building on \citet{Graham:Sahu:22a}, which revealed how dry S0 mergers build E
galaxies --- by folding in disc stars --- Fig.~\ref{Fig-BH_schematic-3} is a
helpful schematic to illustrate how wet mergers involving spiral galaxies can
build dusty S0 galaxies.

Unlike the $M_{\rm bh}$-$M_{\rm *,sph}$ diagram (Fig.~\ref{Fig-MMsph}),
the $M_{\rm bh}$-$M_{\rm *,gal}$ diagram (Fig.~\ref{Fig-MMgal}) offers less
clear relations for the S0 galaxies. Nonetheless, the general offset nature of
the dust-poor and dust-rich S0 galaxies is evident.  Here, the (wet
merger)-built, dust-rich S0 galaxies appear as a high-mass extension to the S
galaxies.  The dust-poor S0 galaxies have notably lower $M_{\rm *,gal}/M_{\rm
  bh}$ ratios for a given black hole mass.  $M_{\rm bh}$-$M_{\rm *,gal}$
relations are provided in the caption to Fig.~\ref{Fig-MMgal}.  The size of
the offset between the different types of disc galaxy (S, dust-poor S0, and
dust-rich S0) are substantial and suggest that considerable gains could be
made by revisiting the ($M_{\rm bh}$-$M_{\rm *,sph}$)- and/or ($M_{\rm
  bh}$-$M_{\rm *,sph}$)-based derivation of the virial $f$-factors used to
convert AGN virial masses into black hole masses \citep{2009ApJ...694L.166B,
  2021ApJ...921...36B}.

The above division between the S0 galaxies is strong but could be better
because of some discrepant systems
This is, however, expected given the partly stochastic nature of mergers and
the changing neighbourhood where galaxies can find themselves. 
This is explored further in Section~\ref{Sec_noncon}.

\subsubsection{Potential dust bias}

As seen in Fig.~\ref{Fig-6panel}, the effects of dust on the adopted
stellar mass-to-light ratio 
(Section~\ref{Sec_MonL}) are insufficient to fully explain the offset between
the S0 galaxies with and without visible signs of dust. However, it could account
for up to half of the offset {\em if} both of the $\sim$0.13 dex offsets (in
$\log\,M_{\rm *,gal}$) mentioned in Section~\ref{Sec_MonL} were to hold.

Of the four previously well-recognised S0 mergers (NGC~1194, NGC~1316,
NGC~5018 and NGC~5128), all appear to have typical $V-$[3.6] colours around 3.6
mag, except for NGC~5018, whose spheroid could be 0.25--0.50 mag too bright at
3.6~$\mu$m \citep[][their Fig.~A1]{Graham:Sahu:22a}.  
Nonetheless, the offset
behaviour in some, if not all, dusty S0 galaxies --- relative to the dust-poor
S0 galaxies --- in the $M_{\rm bh}$-$M_{\rm *,gal}$ diagram appears to be due
to their merger origin noted in Table~\ref{Table-dust}.  This may reflect an
S+S merger, an S+(dust-poor S0) merger, or a gas-rich but initially 
dust-poor S0+S0 merger\footnote{This could occur if some lower-mass S0
  galaxies accreted relatively pristine H{\footnotesize I}/He gas not heavily
  laced with dust.} if it subsequently produces stars that yield a dusty 
interstellar medium.

\subsection{Spiral galaxies}
\label{Sec_Spiral}

The  $M_{\rm bh}$-$M_{\rm *,sph}$ relation for the 25 S galaxies (derived excluding
NGC~1300 and the Milky Way) can be written as 
\begin{equation}
\log(M_{\rm bh}/M_\odot)=
(2.27\pm0.48)[\log(M_{\rm *,sph}/M_\odot)-10.00]+(6.98\pm0.19), 
\label{Eq-yip}
\end{equation}
in good agreement with \citep{2019ApJ...873...85D}.

\begin{figure*}
\begin{center}
\includegraphics[trim=0.0cm 0cm 0.0cm 0cm, width=0.8\textwidth, angle=0]{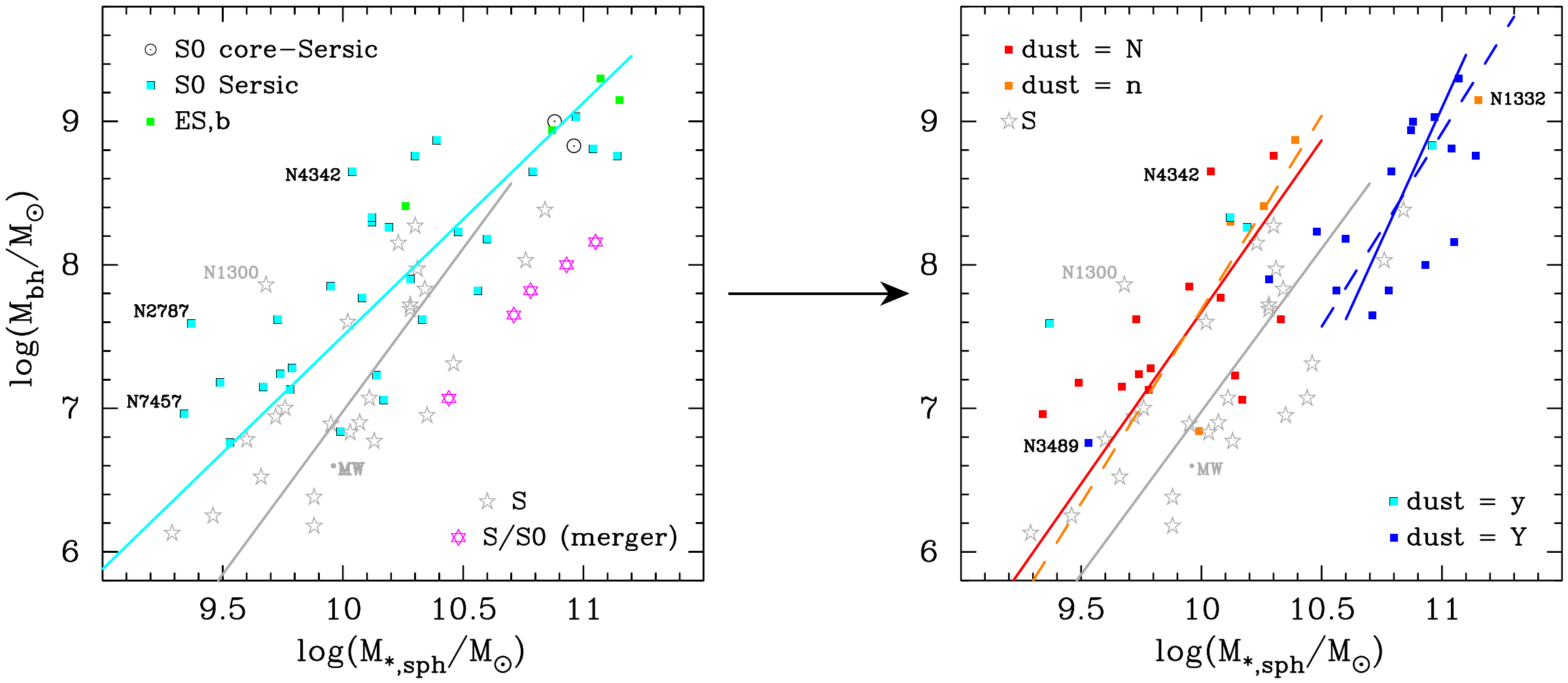}
\caption{Splitting the lenticular galaxies. 
Adaption of the middle panels from Fig.~\ref{Fig-6panel}.
Left: The cyan line, 
$\log(M_{\rm bh}/M_\odot)=
(1.63\pm0.18)[\log(M_{\rm *,sph}/M_\odot)-10.33]+(8.04\pm0.15)$, 
is a fit to the 32 unlabelled S0 galaxies shown
(including two core-S\'ersic galaxies and four ES,b galaxies, but excluding the
previously identified mergers marked in pink, and NGC~404, NGC~2787,
NGC~4342, and NGC~7457).  The grey line, $\log(M_{\rm bh}/M_\odot)=
(2.27\pm0.48)[\log(M_{\rm *,sph}/M_\odot)-10.09]+(7.18\pm0.15)$, 
is a fit to the 25 non-labelled S galaxies (excluding NGC~1300 and the
Milky Way). 
Right: The red line, 
$\log(M_{\rm bh}/M_\odot)=
(2.39\pm0.81)[\log(M_{\rm *,sph}/M_\odot)-9.90]+(7.43\pm0.18)$, 
is a fit to the 13 dust-poor (dust=N) S0
galaxies (red points), excluding the stripped S0 galaxy NGC~4342. 
The dashed orange line is a fit to the 17 red and orange points (dust=N and n) S0
galaxies (excluding NGC~1332 and NGC~4342) and is such that 
$\log(M_{\rm bh}/M_\odot)=
(2.70\pm0.77)[\log(M_{\rm *,sph}/M_\odot)-9.96]+(7.57\pm0.18)$. 
The dashed blue line has the same slope as the dashed orange line but is shifted by an
arbitrary $\log(3.5)\approx0.54$ dex to the right, while the solid blue line
is a poorly defined fit to the 15 blue points, which are the unlabelled dusty
(dust=Y) S0 galaxies, yielding a 
slope of 3.69$\pm$1.51. 
}
\label{Fig-MMsph}
\end{center}
\end{figure*}

\citet{2019ApJ...876..155S} show that the S galaxy $M_{\rm bh}$-$M_{\rm
  *,gal}$ relation predominantly resides to the right of the S0 galaxy $M_{\rm bh}$-$M_{\rm
  *,gal}$ relation.  This can be seen in Figure~\ref{Fig-M-S0-gal} and might be 
explained by ongoing star formation in 
S galaxies while S0 galaxies are, in effect, left behind.
This concept also allows for a certain amount of AGN growth in the S galaxies,
so long as it does not out-pace the stellar growth by too much.
However, 
as we learned in Section~\ref{Sec_S0_split}, this picture is incomplete,
in that the S0 galaxy population is comprised of systems that may have either
been {\em left behind} or advanced to higher stellar masses and
$M_*/M_{\rm bh}$ ratios through major wet mergers.  The S galaxies now appear
as an intermittent population (Fig.~\ref{Fig-MMsph}).  
Those galaxies located to the 
high $M_{\rm *,sph}/M_{\rm bh}$ side of the S galaxy distribution have
experienced substantial mergers.   The S galaxies have probably experienced
less substantial mergers.

There are five S galaxies on this high $M_*/M_{\rm bh}$ envelope with bulges
more massive than $\log(M_{\rm *,sph}/M_\odot) = 10.35$ dex (see
Fig.~\ref{Fig-6panel}).  One of these is the Sa? merger NGC~2960, marked with
a pink hexagon in Fig.~\ref{Fig-M-S0-gal}.  The four other galaxies (NGC~1097,
NGC~1398, NGC~4501, and NGC~2273) are mentioned below.

NGC~1097 is a dusty, barred spiral galaxy undergoing substantial star
formation, likely induced by interaction with its dwarf neighbour ESO~416-19,
aka NGC~1097A \citep{1989ApJ...342...39O}.

NGC~1398 is the brightest galaxy of the NGC~1398 Group (10 members) located
within the Eridanus Cluster associated with the larger Fornax Cluster.  Dust
lanes associated with the bar are evident, as are dust lanes crossing nearly
perpendicular to the bar.  This double-ringed galaxy has the biggest galaxy
stellar mass of all the spiral galaxies in the sample. Despite this, a
cautionary flag is raised as to whether its bulge mass has been over-estimated
due to the potential presence of an (unmodelled) barlens.\footnote{Barlenses,
not to be confused with inner discs, have been associated with bar buckling
events which produce boxy/X/(peanut shell)-shaped `pseudobulges'
\citep{1975IAUS...69..297B, 1975IAUS...69..349H, 1981A&A....96..164C,
  2005MNRAS.358.1477A, 2011MNRAS.418.1452L, 2016MNRAS.459.1276C}.}

NGC~4501 (aka M88) is a dusty star-forming spiral galaxy with a nascent
H{\footnotesize I} tail pointing away from M87 \citep{1990AJ....100..604C} due
to ram-pressure stripping in the Virgo cluster \citep{2008A&A...483...89V}.
The spheroid mass may have been overestimated due to a possible (unmodelled)
upturn in the inner disc light.  A slightly similar situation may exist with
M33 (not included here), which was initially reported to have a bulge
\citep{1979pkdg.conf..271B, 1992AJ....103..104B} that was subsequently
diminished in stature \citep[e.g.,][]{1987AJ.....94..306K,
  1993ApJ...410L..79M}.

Finally, NGC~2273 has a UV-bright star-forming ring structure reminiscent of
that in NGC~2974. Both galaxies have a bulge-to-total stellar mass ratio of
one-third.  Furthermore, NGC~2273 has a nuclear bar and stellar spiral
structure within the main bar \citep{2000ApJS..128..139F,
  2002ApJ...575..814P}, somewhat similar to NGC~2974, with its nuclear gas
spiral within its weak main bar.
NGC~2273 has roughly $10^9$ M$_\odot$ of H{\footnotesize I} and is the
brightest galaxy in the NGC~2273 Group of three, also referred to as the `Lyon
Groups of Galaxies' LGG~137 \citep{1993A&AS..100...47G}.  NGC~2273 is
suspected of having interacted with UGC~03530, aka NGC~2273B
\citep{1987A&A...171...16B}.  In contrast with NGC~2974, the spheroid in
NGC~2273 has an effective half-light radius which is 2.4 times (0.38 dex)
smaller, a velocity dispersion which is 0.22 dex smaller, and a black hole
mass which is 19 times (1.28 dex) smaller.

This subsection's divergence was intended to see if the tabulated spiral
galaxies with reportedly massive spheroids may have had their formation aided
by (expectedly minor)\footnote{Major mergers would destroy the spiral pattern
  and build S0 galaxies.} mergers and interactions.  Because most spiral
galaxies contain dust, with the more massive spirals being dustier
\citep{1990ApJ...364..444V}, alternative signatures of interactions,
accretion, or mergers were searched for in the literature and reported above.
Further research is, however, required on this front, given that many spiral
galaxies, perhaps including those not numbered in Fig.~\ref{Fig-6panel}, may
also show signs of interactions.
Nonetheless, removing NGC~1398 and NGC~4501, the above small sample of (5-2=3)
`spiral' galaxies includes one recognised spiral merger (NGC~2960), the spiral
galaxy NGC~2273 thought to have undergone a past interaction and having
features remarkably similar to that seen in NGC~2974, and the previously
interacting spiral galaxy NGC~1097.  These galaxies are identified in the
lower-middle panel of Fig.~\ref{Fig-6panel}.

Considering S galaxies with spheroid stellar masses below $\log(M_{\rm
  *,sph}/M_\odot) = 10.35$ dex, an additional four spiral galaxies are
labelled in Fig.~\ref{Fig-6panel} because of how they define the continuation
of an apparent high $M_{\rm *,sph}/M_{\rm bh}$ envelope for the S
galaxies.\footnote{Conceivably, absent low surface brightness (LSB) galaxies,
  with large-scale discs may erode this apparent envelope should they contain
  sufficiently massive black holes.}  These include NGC~4826 (aka the Black
Eye Galaxy), with copious amounts of dust and gas counter-rotating to the
stellar disc \citep{1992Natur.360..442B, 1994AJ....107..173R}, likely brought
in by a smaller, gas-rich galaxy.  The others are the Milky Way and NGC~3079,
which are somewhat similar to each other, plus NGC~1320
\citep{2003ApJS..146..353M}, which could perhaps be called the `Eye Liner
Galaxy' with a reasonably prominent dust-arm visible (on just one side of the
galaxy) in the HST/WFPC2/F791W-F547M image available at the HLA.

\begin{figure}
\begin{center}
\includegraphics[trim=0.0cm 0cm 0.0cm 0cm, width=1.0\columnwidth,
  angle=0]{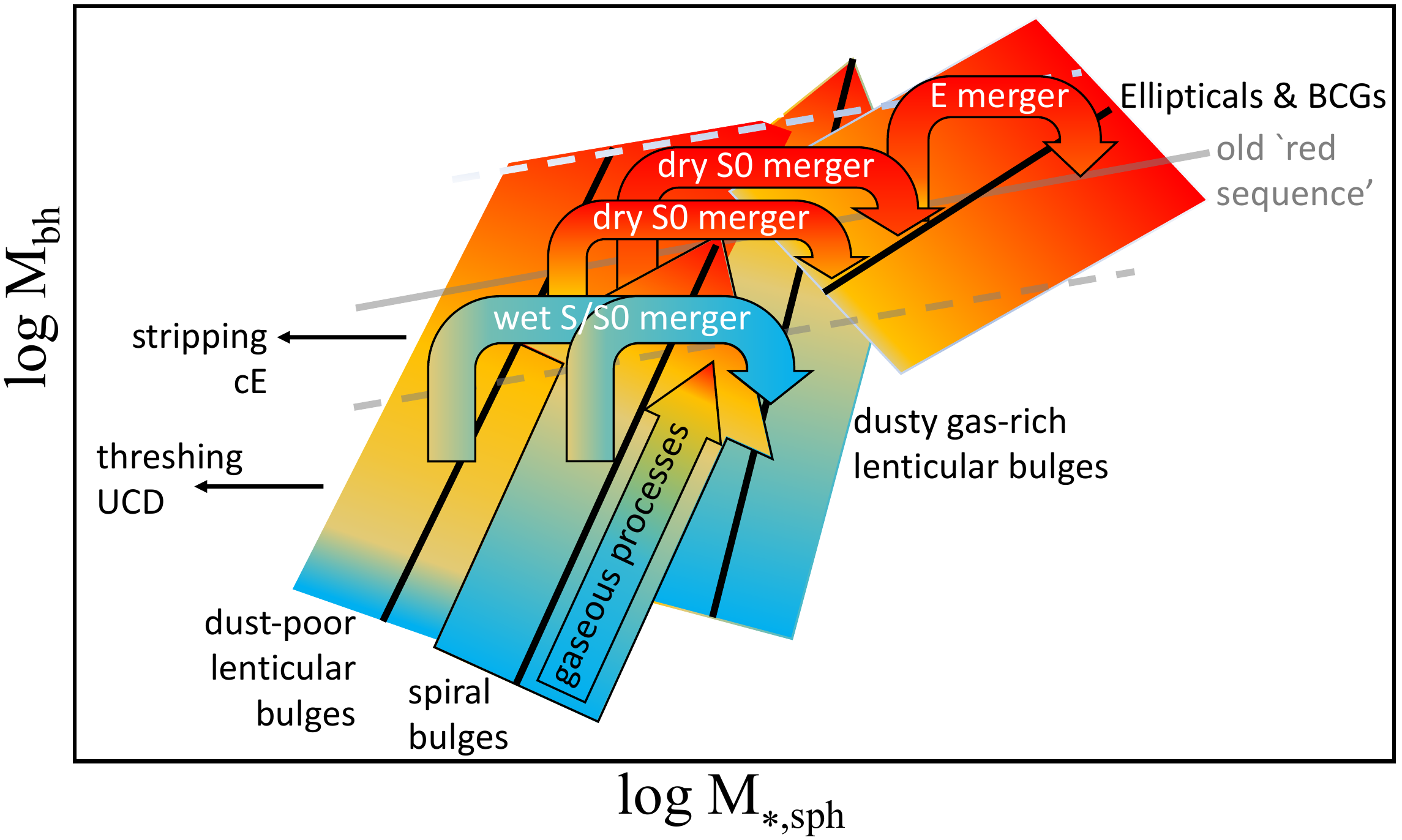}
\caption{Adapted from Fig.~7 in \citet{Graham:Sahu:22b}.  Here, two steep
  $M_{\rm bh}$-$M_{\rm *,sph}$ relations for visibly dust-poor and (wet-merger-built)
 dust-rich lenticular galaxies
  are shown to bracket the steep relation for spiral galaxies (Eq.~\ref{Eq-yip}), which
  previously appeared as an unusually steep relation relative to that of the
  (dust-poor and dust-rich) lenticular galaxies and the (merger-built)
  elliptical galaxies (Fig.~\ref{Fig-M-S0-gal}).
}
\label{Fig-BH_schematic-3}
\end{center}
\end{figure}

\begin{figure*}
\begin{center}
\includegraphics[trim=0.0cm 0cm 0.0cm 0cm, width=0.8\textwidth, angle=0]{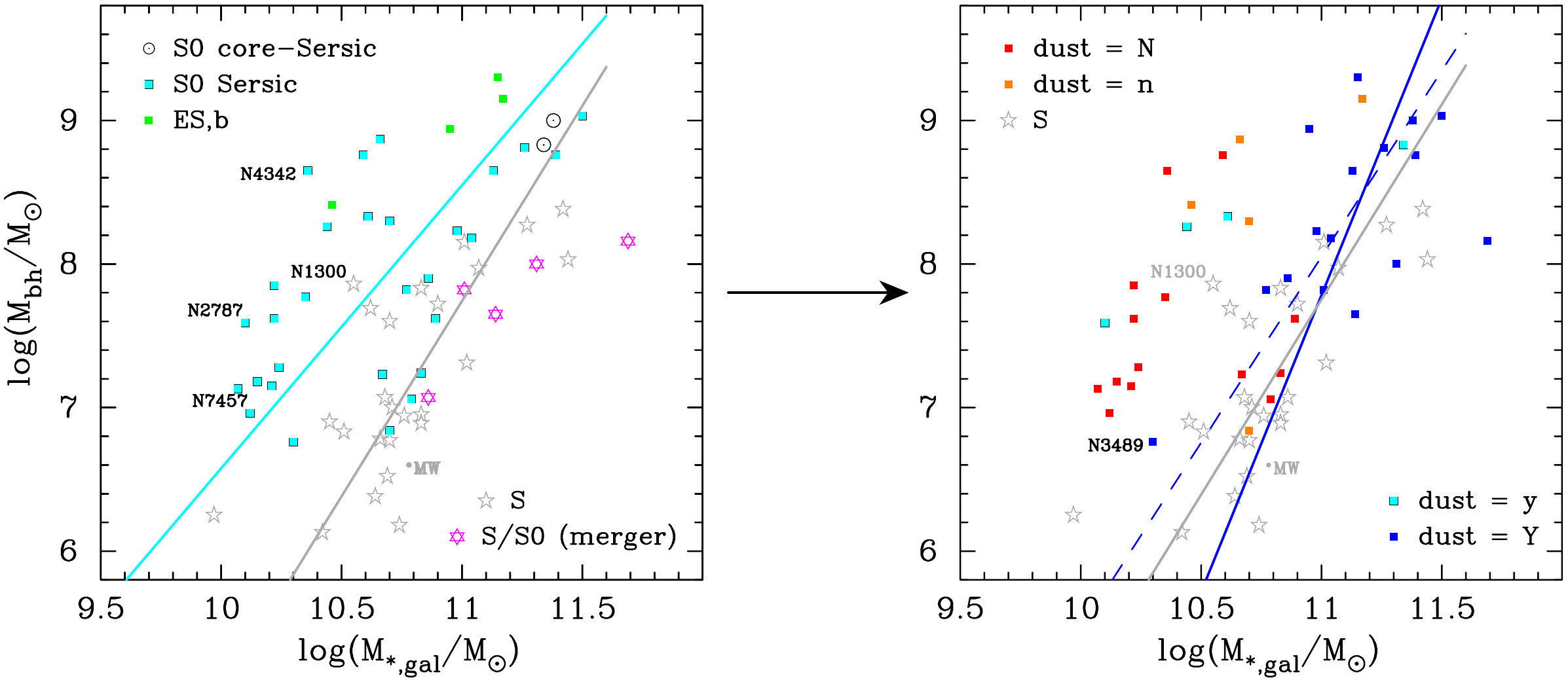}
\caption{Similar to Fig.~\ref{Fig-MMsph} but showing the galaxies' stellar
  masses. This is an adaption of the left panels from Fig.~\ref{Fig-6panel} and
  shows the development achieved when no longer considering S0 galaxies as a
  single population. 
Left: The cyan line, 
$\log(M_{\rm bh}/M_\odot)=
(1.97\pm0.29)[\log(M_{\rm *,gal}/M_\odot)-10.75]+(8.05\pm0.15)$, 
is a fit to the 32 non-labelled S0 galaxies
(including two core-S\'ersic galaxies and four ES,b galaxies but excluding the
five previously identified mergers marked in pink). 
The grey line, 
$\log(M_{\rm bh}/M_\odot)=
(2.72\pm0.49)[\log(M_{\rm *,gal}/M_\odot)-10.79]+(7.18\pm0.14)$, 
is a fit to the 25 non-labelled S galaxies, excluding NGC~1300 and the Milky Way.
Right: The solid blue line excludes NGC~3489 and has an uncertain slope of
4.14$\pm$1.84, while the (biased) dashed blue line includes NGC~3489 and is
such that $\log(M_{\rm bh}/M_\odot)= (2.59\pm0.54)[\log(M_{\rm
    *,gal}/M_\odot)-11.09]+(8.28\pm0.18)$.  They represent fits to the S0
galaxies from the `Y' dust bin, denoted by the dark blue (not cyan) symbols.
The present sample of dust-poor S0 galaxies does not have a clear $M_{\rm
  bh}$-$M_{\rm *,gal}$ relation.  }
\label{Fig-MMgal}
\end{center}
\end{figure*}

\subsection{Interesting and/or nonconforming galaxies}\label{Sec_noncon}

As noted in Section~\ref{Sec_Excl}, looking for galaxies which do not conform
with the general trend may be insightful.  While some systems might represent
measurement error, others may reflect true scatter and thus further insight
into the evolutionary paths of galaxies.  An attempt has been made here to
provide additional information on interesting or nonconforming galaxies if
not already discussed in Section~\ref{Sec_data}. 
This subsection is considered additional to the main result and can be bypassed
by readers not after every detail. 

There is a cluster of five dust-poor S0 galaxies (NGC~1023, NGC~3384,
NGC~4371, NGC~4762, and NGC~7332) in the lower-right of the distribution 
in the $M_{\rm bh}$-$M_{\rm *,gal}$ diagram (Fig.~\ref{Fig-MMgal}, right-hand
side).  They overlap with the spiral galaxies and the low-mass extrapolation of the
$M_{\rm bh}$-$M_{\rm *,gal}$ relation for dust-rich S0 galaxies. 
While they are not major merger remnants, four of these five show signs of 
accretion or minor mergers.
NGC~1023 and NGC~7332 have chemically decoupled stellar nuclei \citep{1999AJ....117.2725S}.
NGC~7332 has a 2.5$\pm$0.5 Gyr old nucleus, a gas disc that is counter-rotating with respect to the
stellar disc \citep{1994AJ....107..160F} and a small H{\footnotesize I} mass
of $4\times10^6$ M$_\odot$ \citep{2012MNRAS.422.1835S}. 
It is unclear if the nuclear disc was solely built by accretion or if the bar
had a hand to play \citep{2014MNRAS.445.3352C}. 
NGC~1023 has a 7~Gyr old nucleus within an older bulge and a reasonably thin/flat stellar
disc.  It has a disturbed large-scale H{\footnotesize I} disc 
\citep{1979MNRAS.187..537A, 1984MNRAS.210..497S} of $2\times10^9$ M$_\odot$
\citep{2012MNRAS.422.1835S}, giving $M_{H{\footnotesize I}}/M_{\rm *,gal}=0.026$, 
plus outer kinematics \citep{2008MNRAS.384..943N, 2012MNRAS.423.2957B} 
and a nuclear, stellar disc \citep{2014ASPC..486..141D} suggestive of a past
interaction or small merger event. 
Despite the H{\footnotesize I} gas mass, no dust is visible, 
presumably still diffusely distributed or tied up in the gas phase metallicity. 

Another galaxy in the grouping of five is NGC~3384, which may be interacting
with a supergiant intergalactic H{\footnotesize I} ring in the Leo I Group
\citep{2003ApJ...591..185S}.  This galaxy contains little noticeable dust in the
optical images \citep{2000AJ....120..123T} but has been detected in CO (J=2-1
transition), yielding an H$_2$ mass of $(3.5\pm0.8)\times10^6$ M$_\odot$
\citep{2003ApJ...584..260W}, and it has a global H{\footnotesize I} mass of
$\sim$2$\times10^7$ M$_\odot$ \citep{2012MNRAS.422.1835S}.
NGC~4371, with the smallest black hole mass of the five, stands out for not
displaying any obvious signs of past/ongoing interaction, accretion, or minor
merger activity, as seen in the other four (growing) galaxies.
However, it has a second nuclear star-forming ring with a major-axis
radius of $\sim$10$\arcsec$ ($\sim$0.8 kpc) \citep{2010MNRAS.402.2462C}. 
Finally, NGC~4762 has a thick asymmetric gas disc and mild asymmetry in its stellar structure
\citep{1984ApJS...56..283W, 1994AnA...286L...5W}.  It has a warped S-shaped
outer stellar disc and a blue lens possibly built from star formation
associated with a neighbourly interaction or the consumption of a smaller
galaxy \citep{1984ApJS...56..283W, 1995ApLnC..31..165W}.  In passing, it is
noted that NGC~3115 (in sample) and NGC~5866 (not in sample) may represent
more dusty analogues of NGC~4762, having more substantial thick stellar discs 
\citep{1979ApJ...234..829B}.  The
dusty S0 galaxy NGC~5252, another known merger, represents an even more
evolved state of affairs, in which the stellar disc has thickened
considerably, 
but the edge-on thin disc is still apparent.  Curiously, NGC~4762 has a rather
low spheroid-to-galaxy stellar mass ratio for a merger, at just 0.08, but
perhaps this is reflective of what some minor mergers produce. 

In addition to NGC~3115, mentioned above, 
there is another dusty ES,b galaxy in the sample.  It is NGC~6861, 
shown in Fig.~\ref{Fig-dust}, and 
dubbed here `The Speaker' due to its strong dust rings over the inner
kpc, which make it resemble an audio speaker.
NGC~6861 is interacting with NGC~6868 in the Telescopium
galaxy group \citep{2010ApJ...711.1316M}. 
It is interesting because it is a reasonably 
compact massive galaxy ($R_{\rm e,gal}\sim2.5$ kpc,
$M_{\rm *,gal}\sim 10^{11}$ M$_\odot$) reminiscent of relic red nuggets. 
Nevertheless, it also appears to have grown/accreted an intermediate-scale
disc suggestive of a wet merger event that {\em might} have contributed to its
spheroid's development. 

Also near the top of the 
$M_{\rm bh}$-$M_{\rm *,gal}$ relation for dust-rich S0 galaxies
(Fig.~\ref{Fig-MMgal}) 
are the two core-S\'ersic S0 galaxies: NGC~524 and NGC~5813 
\citep{2009MNRAS.399.1839K, 2011MNRAS.415.2158R, 2014MNRAS.444.2700D,
  2015MNRAS.452....2K}.
Core-S\'ersic galaxies are thought to have been built from
mergers in which a binary massive black hole scoured away the central stellar 
phase-space of what may have been a \citet{1963BAAA....6...41S} $R^{1/n}$
light profile \citep{1980Natur.287..307B, 1991Natur.354..212E,
  2004ApJ...613L..33G, 2013degn.book.....M}.  Additional core-S\'ersic S0 galaxies
can be found in \citet{2013ApJ...768...36D}.  As mentioned in
Section~\ref{Sec_Intro}, the core-S\'ersic galaxy 
NGC~5813 is immersed in a group-sized X-ray halo which has shut down
star formation.  While NGC~5813 is an old merger \citep{1995A&A...296..633H},
it is less clear when NGC~524 experienced its merger event. 
NGC~524 is immersed in a smaller, galaxy-sized X-ray halo
\citep{2004MNRAS.350.1511O}  
that has not yet removed all the dust and cold gas from the
centre of NGC~524 \citep{2000AJ....120..741S}.  
These two core-S\'ersic S0 galaxies are interesting in that 
depleted cores are generally considered to be a sign of dry mergers
\citep[e.g.][]{2007ApJ...671...53M, 2008ApJ...686..432S} 
due to the efficient gravitational drag on black holes from gas clouds and/or
a circumbinary disc sparing the ejection
of hypervelocity stars from the core region of a wet merger 
\citep{2006ApJ...648..976M, 2006ApJ...651..392S, 2007Sci...316.1874M,
  2017MNRAS.464.3131K}.  However, the dust 
in these two galaxies reveals a more complex history, affecting 
accretion and feedback \citep{2022arXiv221111788L}. 

Returning to the low-mass end, another system which stood out is the S0 galaxy 
NGC~7457, already mentioned in Section~\ref{Sec_prev} as a merger remnant 
revealed through its kinematics.  
Apart from a faint non-nuclear dust ring, it appears dust-poor. 
It has been reported to have an H$_2$ mass of 
$(3.3\pm1.0)\times10^6$ M$_{\odot}$, comparable to the $5.5\times10^6$
M$_{\odot}$ of H$_2$ gas reported by 
\citet{2003ApJ...584..260W} in the gas-rich dwarf 
galaxy NGC~404, which has an order of magnitude less stellar mass. 
Like NGC~404, NGC~7457 resides 
in the lower-left of the $M_{\rm bh}$-$M_{\rm *,gal}$ diagram.  NGC~7457 has a
young $\sim$2 Gyr old nucleus and experienced star-formation 2-3
Gyr ago, giving rise to new globular clusters \citep{2008AJ....136..234C} and
perhaps the now faint dust ring. 
However, it is a dust-poor post-merger residing among the low-mass S0 galaxies.
Although not S0 galaxies, 
NGC~3377 (ES,e) and Circinus (S), mentioned in Section~\ref{Sec_data}, also
appear to be post-merger systems residing among the low-mass S0 galaxies.
These three galaxies are enclosed with black squares in Fig.~\ref{Fig-M-S0-gal}. 

\subsection{Updating the typical $B/T$ ratio in dry S0 mergers}
\label{Sec_BonT}

Unlike  wet mergers, where the progenitor galaxies' $B/T$ stellar mass
ratios are harder to establish given the likelihood of star formation during
the merger, this is not a problem for dry mergers.  The dry
major-merger-induced-jump shown by \citet[][section~3.2, final
  paragraph]{Graham-sigma}, to convert two equal S0 galaxies into an E galaxy,
was based on an $M_{\rm bh}$-$M_{\rm *,sph}$ relation defined by all of the S0
galaxies.  This jump from the S0 galaxy $M_{\rm bh}$-$M_{\rm *,sph}$ relation
to the E galaxy $M_{\rm bh}$-$M_{\rm *,sph}$ relation was shown to equate to
the merger of two S0 galaxies with a bulge-to-total stellar mass ratio of
0.37.  Removing the offset dust-rich S0 galaxies and focussing on the
dust-poor S0 galaxies, they define a steeper relation in the $M_{\rm
  bh}$-$M_{\rm *,sph}$ diagram, which can be seen offset to smaller $M_{\rm
  *,sph}$ at $M_{\rm bh} \sim 10^8$ M$_{\odot}$ (Fig.~\ref{Fig-MMsph}).  As a
consequence, a dry merger induced jump from the new dust-poor S0 galaxy
$M_{\rm bh}$-$M_{\rm *,sph}$ relation to the unchanged E galaxy $M_{\rm
  bh}$-$M_{\rm *,sph}$ relation is still associated with the same doubling of
the black hole mass. However, it now requires a greater increase in the spheroid mass.
This means that more disc stars from the two progenitor galaxies are
required to create the new, bigger spheroid, i.e., the elliptical
galaxy.  The merger of two S0 galaxies with $\log(M_{\rm bh}/M_\odot)=7.7$
dex, and $\log(M_{\rm *,sph}/M_\odot)=10.01$ dex, on the dust-poor S0 galaxy
$M_{\rm bh}$-$M_{\rm *,sph}$ relation (shown in the right-hand panel of
Fig.~\ref{Fig-MMsph}) required to build an E galaxy with $\log(M_{\rm
  bh}/M_\odot)=8.0$ dex on the E galaxy $M_{\rm bh}$-$M_{\rm *,sph}$ relation
(Fig.~\ref{Fig-M-S0-gal}), requires an increase in spheroid mass of 0.84 dex.
This balance is achieved by folding in the disc and bulge stars of two S0 galaxies
with $B/T=0.29$.  This ratio is in line with observations of S0
galaxy $B/T$ ratios \citep{2008MNRAS.388.1708G}.

\subsection{The $M_{\rm *,sph}$-$R_{\rm e,sph}$ diagram}
\label{Sec-size-mass}

\citet{2022MNRAS.514.3410H} recently reported that some local spheroids with
$10.0 < \log(M_{\rm *,sph}/M_\odot) < 10.6$ dex may be the relics of `red
nuggets' observed at $z\sim 2$ \citep{2005ApJ...626..680D,
  2011ApJ...739L..44D}.  They have comparable sizes, masses, and volume number
density.

\begin{figure}
\begin{center}
\includegraphics[trim=0.0cm 0cm 0.0cm 0cm, width=1.0\columnwidth,
  angle=0]{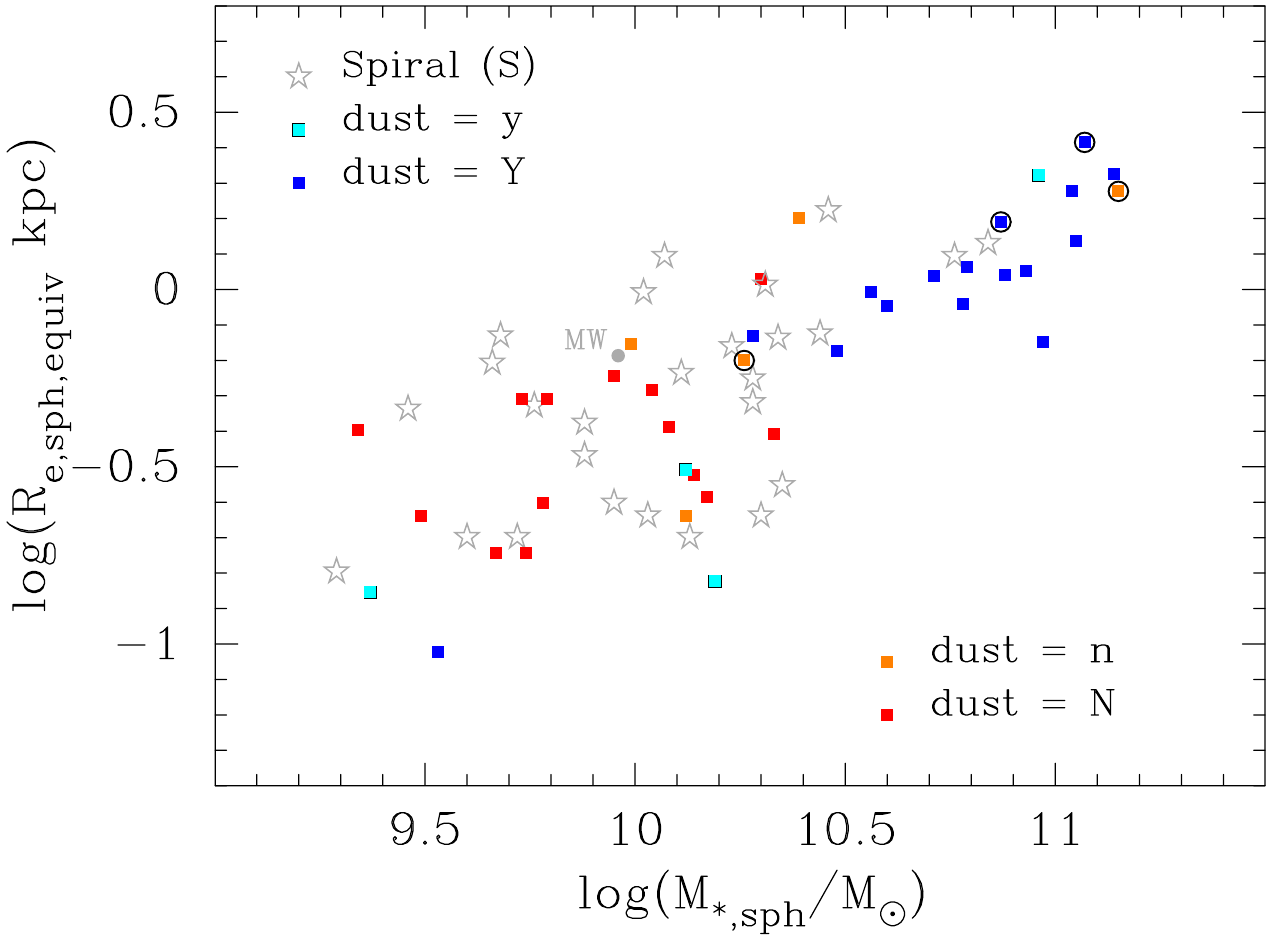}
\caption{Size-mass diagram for spheroids in disc galaxies.  The four ES,b
  galaxies are circled. 
}
\label{Fig-size-mass}
\end{center}
\end{figure}

Figure~\ref{Fig-size-mass} shows the size-(stellar mass) diagram for the
bulges of the disc galaxies studied here.  The ES,b systems such as NGC~6861
--- which are bulge-dominated and possibly largely unevolved relics from
$z\sim2.5$ --- have the same sizes and stellar masses as the dusty S0 galaxy
bulges thought to have been built/bolstered by wet mergers at lower
redshifts. This is not necessarily a problem.  The $M_{\rm *,sph}$-$R_{\rm
  e,sph}$ trend seen in Fig.~\ref{Fig-size-mass} and explored in
\citet{Graham-sigma} and \citet{HGS2022} is expected given the $M_{\rm
  *,sph}$-$\sigma$ relation for S0 galaxies, coupled with the virial theorem
approximation $M_{\rm *,sph} \propto \sigma^2R_{\rm e,sph}$.
Fig.~\ref{Fig-size-mass} reveals that this seems to hold irrespective of the
presence of an intermediate-scale disc (ES,b galaxy) or a large-scale disc (S0
galaxy).

Of note here is that one may have a compact massive spheroid, with, say,
$\log(M_{\rm *,sph}/M_\odot) >$ 10.6--11 dex and $R_{\rm e,sph} >$ 1--2 kpc,
that was recently ($z <$ 1--2) built by a wet merger.  Such a spheroid is not
a `relic' from the early Universe.  This is not to say that ES,b galaxies
which accreted an intermediate-scale disc are not relics, only that the
evolutionary pool appears to be muddied at $10.6 \lesssim \log(M_{\rm
  *,sph}/M_\odot) \lesssim 11.2$ dex by major wet mergers that build new
spheroids in dusty S0 galaxies.  Presumably, from Fig.~\ref{Fig-MMgal}, the
compact bulges of the S0 galaxies in \citet{2022MNRAS.514.3410H}, with $10 <
\log(M_{\rm *,sph}/M_\odot) < 10.6$ dex, were found in dust-poor galaxies.  As
seen in Fig.~\ref{Fig-M-S0-gal}, the spheroidal component of dust-poor ES,e
and true E galaxies tend to have distinctly bigger sizes ($R_{\rm e,sph} > 2$
kpc).  As such, they would not have been counted among the potential relic red
nuggets. Those systems do not contain a preserved nugget but have been
transformed by major mergers \citep{Graham:Sahu:22a}.

\section{Discussion}\label{Sec_Disc}

\subsection{Mergers: shaking the dominance of AGN feedback}

After reviewing the literature, it is apparent that the dusty S0
galaxies are known merger products, while the relatively dust-poor  S0 galaxies
are not recognised as such --- although some are reported to have experienced
minor accretion or disturbances (Table~\ref{Table-dust}).
Fig.~\ref{Fig-BH_schematic-3} qualitatively illustrates how wet (gas-rich) 
mergers evolve the dust-poor
S0 galaxies across to the dust-rich S0 galaxy sequence in the $M_{\rm
  bh}$-$M_{\rm *,sph}$ diagram, quantitatively shown in Fig.~\ref{Fig-MMsph} but
without the evolutionary tracks. 
For a quarter of a century, AGN feedback has been heralded as the
driving force behind the black hole scaling relations
\citep[e.g.][]{1998A&A...331L...1S, 2012ARA&A..50..455F}.  However, 
mergers also have a considerable yet often understated 
role. Indeed, mergers have a dominant role in creating the E (and ES,e)
galaxies, for which the bulk of the progenitor's disc stars are folded into
the E galaxy \citep{Graham:Sahu:22a, Graham-sigma}.

Collisions aside, stars tend to form in discs 
rather than in bulges. 
Therefore, to fully understand the $M_{\rm bh}$-$M_{\rm *,sph}$ relations requires 
an element in which some disc stars eventually become bulge stars. 
Gas-fuelled AGN and disc stellar growth need not be automatically accompanied
by bulge stellar growth.  
Indeed, if AGN feedback regulated the star formation,
one might expect to see stronger 
$M_{\rm bh}$-$M_{\rm *,disc}$ and $M_{\rm bh}$-$M_{\rm *,gal}$
relations than the $M_{\rm bh}$-$M_{\rm *,sph}$ relations. 
Furthermore, 
quasar feedback is not isotropic but occurs in bi-directional polar jets, limiting
its ability to quench star formation within the disc plane. It is, therefore,
perhaps not surprising that mergers, moving disc stars into bulges,
is an important ingredient of galaxy/black hole evolution.

\subsection{Some areas of impact and predictions}

\subsubsection{Black hole mass functions}

The ability to make refined predictions of the central black hole mass in
other galaxies benefits various research programmes.  This includes work on the
local black hole mass function \citep{1999MNRAS.307..637S,
  2009MNRAS.400.1451V, 2013CQGra..30x4001S} and the masses of active black
holes in distant quasars \citep{2009ApJ...699..800V, 2010AJ....140..546W}.
Initially, a single black hole scaling relation tended to be applied to all
galaxies, irrespective of their morphology.  Knowing the galaxy or spheroid
luminosity function of, say, the LTGs and the ETGs, or their distribution of
S\'ersic indices or velocity dispersions, one could construct the LTG and ETG
black hole mass function \citep[e.g.][]{2009MNRAS.400.1451V}.  However, one
can now do better than this, given that the black hole mass is related in
different ways to the spheroid stellar mass of ETGs of different morphology,
reflecting their merger history.  For a given spheroid stellar mass, the black
hole mass of an ETG can differ by an order of magnitude or more.  This
adjustment will considerably impact the black hole mass function, especially
at low masses, due to the steepness of the $M_{\rm bh}$-$M_{\rm *,sph}$
relations.

\subsubsection{AGN masses}

\citet{2015AJ....149..155K} present eight dusty Seyfert~2 S0 and S galaxies,
including the dusty S0 galaxy 
NGC~5252, with its suspected binary SMBH \citep{2015ApJ...814....8K, 2017MNRAS.464L..70Y}.  
These eight galaxies are merger remnants showing a fading AGN.  
Other examples of galaxies with a directly measured black hole mass but with 
possible evidence for a binary SMBH are the spiral galaxy 
NGC~4151 \citep{2012ApJ...759..118B} and the dusty S0 galaxy NGC~1194
\citep{2015KPCB...31...13V, 2016AN....337...96F}. 
It seems reasonable 
to speculate that, if built up by mergers, these eight Seyfert galaxies 
should have black hole masses, which place them on the right-hand side of the S0
galaxy $M_{\rm bh}$-$M_{\rm *,gal}$ relation (Fig.~\ref{Fig-MMgal}, left-hand side). That is, they
are predicted here to have $M_{\rm bh}/M_{\rm *,gal}$ ratios an order of
magnitude smaller than typically observed in dust-poor S0 galaxies, as is observed with 
NGC~1194, NGC~4151, and NGC~5252.  This has nothing to do with pseudobulges
built from secular evolution; quite the opposite, it is because their bulges
have been built, in part, by mergers which folded in some pre-existing disc 
stars and thereby lowered the $M_{\rm bh}/M_{\rm *,sph}$ ratio.  
The dust-rich S0 galaxy $M_{\rm bh}$-$M_{\rm *,gal}$ relation, 
which somewhat overlaps with the S galaxy $M_{\rm bh}$-$M_{\rm *,gal}$ relation
(Fig.~\ref{Fig-MMgal}, right-hand side), has been used to predict these eight
galaxies' central black hole mass (Table~\ref{Table-Seyfert}).

\begin{table}
\centering
\caption{Predicted $M_{\rm bh}$ for Seyfert galaxies from \citet{2015AJ....149..155K}.}\label{Table-Seyfert}
\begin{tabular}{lrccc}
\hline
Seyfert galaxy      &  $K_s$ (mag) &  $D_L$ (Mpc) &  $\log(M_{\rm *,gal}/M_\odot)$ &  $\log(M_{\rm bh}/M_\odot)$ \\
\hline
Mkn 1498                  &  11.144  &   234  &  11.6  &  9.6  \\
NGC 5252                  &   9.768  &   104  &  11.4  &  9.2  \\
NGC 5972                  &  10.688  &   126  &  11.2  &  8.7  \\
SDSS 1510+07              &  12.862  &   197  &  10.8  &  7.4  \\
SDSS 2201+11              &  10.831  &   120  &  11.1  &  8.4  \\
SDSS 1430+13              &  12.751  &   375  &  11.4  &  9.0  \\
UGC 7342                  &  12.212  &   207  &  11.1  &  8.2  \\
UGC 11185                 &  ...     &   ...  &  ...   &  ...  \\
\hline
\end{tabular}

Column~1: Galaxy name: 
SDSS 1510+07 = SDSS J151004.01+074037.1; 
SDSS 2201+11 = SDSS J151004.01+074037.1; 
 = SDSS J143029.88+133912.0 (aka The Teacup). 
Column~2: 2MASS $K_s$-band galaxy magnitude available in NED.
The AGN-to-bulge luminosity is only a few per cent, and thus no effort has been
made to subtract the AGN flux.
Column~3: Luminosity distances based on a Hubble constant of 73 km s$^{-1}$
Mpc$^{-1}$, $\Lambda_m=0.3$, and $\Lambda_\Omega=0.7$, taken from NED. 
Column~4: Estimated galaxy stellar mass based on a constant $K_s$-band 
$M/L$ ratio of 1.0. 
Column~5:  Predicted black hole mass based on the dust-rich S0 galaxy 
$M_{\rm bh}$-$M_{\rm *,gal}$ relation given in Fig.~\ref{Fig-MMgal}. 
\end{table}

\subsubsection{Gravitational waves}

The benefits of refined black hole mass predictions from (galaxy
morphology)-dependent black hole scaling relations can also be extended to efforts to
predict the gravitational wave background from coalescing massive black hole
binaries \citep[][]{2005LRR.....8....8M, 2012ApJ...744...74G,
  2013MNRAS.433L...1S, 2018MNRAS.474.5672L, 2021MNRAS.502L..99M}, plus
gravitational wave signals
from stellar mass black holes and neutron stars as they merge with massive
black holes in what is known as an `extreme mass ratio inspiral' (EMRI) event
\citep{2001CQGra..18.4067H, 2012A&A...542A.102M, 2017PhRvD..95j3012B}.  These
refinements are of importance for 
planned detectors such as the Laser Interferometer Space Antenna
({\it LISA})\footnote{\url{www.lisamission.org}} \citep{1997CQGra..14.1399D,
  2022arXiv220306016A} and the Chinese TianQin mission \citep{2016CQGra..33c5010L,
  2021PTEP.2021eA107M} aiming to detect gravitational waves in the
sub-milliHertz to 1 Hz range.  The refined relations should also have importance for current
endeavours to detect the gravitational wave background arising from
numerous coalescing massive black holes and monochromatic signals from
individual binaries.  By better populating (mock or real) catalogues of galaxies
with more accurate black hole masses and with a better knowledge of which
(real) galaxy 
types experienced a major merger, it enables improved predictions of
massive black hole binaries and thus an improved knowledge of the collective
ocean of gravitational waves that they generate. This, in turn, benefits our
expectations for the local
influence of these waves on the arrival time of pulsed emission from pulsars located in an
array around us \citep{1990ApJ...361..300F, 2008MNRAS.390..192S,
  2019A&ARv..27....5B}.  Ongoing experiments to indirectly detect these waves include the Parkes
Pulsar Timing Array 
\citep[PPTA:][]{2013PASA...30...17M, 2013Sci...342..334S, 2022ApJ...932L..22G}, 
the European Pulsar Timing Array
\citep[EPTA:][]{2015MNRAS.453.2576L, 2021MNRAS.508.4970C}, the North American
Nanohertz Observatory for Gravitational Waves
\citep[NANOGrav:][]{2021ApJ...914..121A, 2021MNRAS.502L..99M}, plus newcomers
such as the Indian Pulsar Timing Array \citep{2022PASA...39...53T}, and the South
African MeerTime Pulsar Timing Array \citep{2022PASA...39...27S}.  It is thus
a large global enterprise.  The new (galaxy
morphology)-dependent black hole scaling relations, with their order of
magnitude differences in $M_{\rm bh}/M_{*,sph}$ ratio, is valuable knowledge
in one's toolkit.  Knowing which galaxies have formed from major
merger events, which were gas-poor or gas-rich
mergers (impacting the binary BH merger timescale), what progenitor galaxies
likely built the merger product (giving insight on the initial BH masses), and what the expected
final black hole mass is based on the galaxy morphology, offers significant
advantages over the past use of a single near-linear `red sequence'.  These
advances 
should help with the interpretation, and perhaps even the discovery, of
nanoHertz to 1 Hertz gravitational waves.

\subsubsection{Evolution of the scaling relations}

Studies of possible evolution in the $M_{\rm bh}$-$M_{\rm *,sph}$ and $M_{\rm
  bh}$-$M_{\rm *,gal}$ relations with look back time to higher redshifts
\citep[e.g.][]{2007ApJ...667..117T, 2010ApJ...708..137M, 2015ApJ...802...14S}
will also benefit from a greater awareness of (galaxy morphology)-dependent
relations. 
For instance, the scaling relation for quasars in distant disc galaxies is
likely to be different from the relation defined by a local sample dominated
by E galaxies.  A measured offset in these two relations
(high-z AGN in disc galaxies and low-z E galaxies) may reflect the
apple versus orange sample selection.  Indeed, the local S and E galaxy
$M_{\rm bh}$-$M_{\rm *,sph}$ 
relations differ. This is obviously not due to evolution over different
look-back times because both samples have the same cosmological time stamp.  As
the local benchmark is now better understood, with a suite of (galaxy
morphology)-dependent scaling relations, and as it becomes better
calibrated over time, improvements in measurements of the evolution in the
scaling relations can be made.  Due to its spatial resolution, aperture size,
and infrared capabilities, the James Webb Space Telescope 
\citep[{\it JWST}][]{2006SSRv..123..485G} is expected to aid such endeavours.

\subsubsection{Tidal disruption events}

Better estimates can also be made for the masses of black holes which induce
flares from their nearby stars, torn open by the gravitational differential across
them as they approach the event horizon and experience a `tidal disruption
event' \citep[TDE:][]{2017MNRAS.465.3840C, 2020NatCo..11.5876S}, which fuels
an accretion disc.  For example, the extragalactic transient Swift
J164449.3+573451 (aka Sw J1644+57) is a TDE associated with a black hole
estimated to have a mass $<2\times10^7$ M$_\odot$ \citep{2011Sci...333..203B, 2011Sci...333..199L}.
The appendix of \citet{2011Natur.476..421B} notes that this TDE occurred at
the centre of a spiral
galaxy and that the galaxy luminosity is $\log(L_H/L_{\odot,H}) = 9.58$ dex.
Adopting even a high $H$-band $M/L$ ratio of 2, the spiral galaxy $M_{\rm 
  bh}$-$M_{\rm *,gal}$ relation seen in Fig.~\ref{Fig-MMgal} predicts the discovery
(not previously announced) 
of an intermediate-mass black hole (IMBH) with $M_{\rm bh}=0.5\times10^5$ M$_\odot$.
This is more than two orders of magnitude smaller than the published upper limit 
obtained using the old near-linear `red sequence' for bulges.  
For comparison with another IMBH in a spiral galaxy,  
it is noted that LEDA~87300 has a galaxy stellar mass of $2.4\times10^9$
M$_\odot$, 
and its central AGN has an estimated black hole mass of 
$0.3\times10^5$ M$_\odot$ \citep{2015ApJ...809L..14B, 2016ApJ...818..172G}.

\subsubsection{S0 galaxies at low masses}

It would be interesting to examine dust-poor S0 galaxies with $M_{\rm *,gal}
\lesssim 10^{10}$ M$_\odot$ ($M_{\rm *,sph}
\lesssim 0.3\times10^{10}$ M$_\odot$).  Figure~\ref{Fig-6panel} suggests that
they may have $M_{\rm bh} \lesssim 10^7$ M$_\odot$, which is currently 
restricted mainly to the purview of the S galaxies.  Dwarf S galaxies are, however,
rare, while dwarf S0 galaxies are abundant. 
It may therefore be that many IMBHs are waiting to be
discovered among the low-mass dwarf S0 galaxies.  Some will inevitably be the
delivery vehicle for massive black holes into S galaxies
\cite[e.g.][]{2021ApJ...923..146G}. 

Given the merger-driven evolution to higher masses across the 
$M_{\rm bh}$-$M_{\rm *,sph}$ diagram, 
the future creation of dusty S0 galaxies with $M_{\rm
  bh}\approx 10^6$ M$_\odot$ and $M_{\rm sph}\approx 10^{10}$ M$_\odot$ is
also predicted here.  Such
S0 galaxies, with $M_{\rm bh}/M_{\rm sph} \sim 10^{-4}$, are absent in the
current data set.  however, they are an anticipated member of the galaxy population
--- unless 
the increasing rate of cosmological expansion prevents their formation
\citep{2013JCAP...04..010E}.

\subsubsection{Specific SFRs and SFHs across the $M_{\rm bh}$-$M_{\rm *,sph}$ diagram}

Coupled with an awareness that the
elliptical galaxies built from dry mergers are offset to the right of the dust-poor S0
galaxies in the $M_{\rm bh}$-$M_{\rm *,gal}$ diagram, it is noted here that
a star formation history (SFH)  gradient likely exist across the face of the $M_{\rm bh}$-$M_{\rm *,gal}$
diagram, from the dust-poor S0 galaxies to dusty S and S0 galaxies, and then
on to the E galaxies \citep{Graham:Sahu:22a}.  An 
additional mass-dependent trend of drying-out and reduced specific
star-formation rate (SFR), and SFH, may also exist along the currently observed S and
dusty S0 galaxy
$M_{\rm bh}$-$M_{\rm *,gal}$ relations as things proceed to higher  masses (Fig.~\ref{Fig-MMgal}). 
While showing these trends would not be a discovery, it would be a value-added
figure explained by the underlying distribution of galaxy morphology.

\subsection{Environment}

The presence or absence of cold gas, and a galaxy's ability to attain cold gas, can
gravely affect its evolution.  
Many mechanisms can remove or prevent the presence of cold gas and thus 
curtail star formation and 
quasar activity, effectively halting a galaxy in its evolutionary track
across the galaxy/black hole scaling diagrams. 

In LTGs --- used here as a reference given their dusty nature --- 
metals in the gas phase of their interstellar medium (ISM) 
can amount to a few (2--4) times the dust mass \citep{2022A&A...668A.130C}.  As the
gas cools in the dense ISM, these metals adhere to and grow mantles
around the refractory dust cores/grains produced during star formation.  
This build-up of the dust helps with the formation of molecular gas clouds by
shielding them from ultraviolet rays capable of dissociating molecules like
H$_2$ and C0, which are both a signpost of dust \citep{2013MNRAS.432.1796A,
  2017A&A...605A..74K, 2019A&A...622A..87K} and stepping stone toward the
gravitational collapse of the gas clouds to form stars.  
Therefore, retaining dusty
gas --- comprised of graphite, silicon carbide, aluminium oxide and other
molecules, which condensed out of the cooling winds of AGB stars or the
decompressing gas of supernova \citep[e.g.][]{2004ARA&A..42...39C} --- can
help raise a further generation of stars. 

While falling into a galaxy cluster can result in the removal of gas and prevent the
acquisition of new gas \citep{1972ApJ...176....1G, 1980ApJ...237..692L}, 
 quasi-isolation or membership of a small galaxy group may result in
the ongoing accretion of gas clouds which replenish this fuel supply without
destroying a galaxy's disc \citep{2016ASSL..420..191F}.  Small galaxy groups
are also more conducive than clusters to galaxy collisions.  These collisions 
can shock and compress gas 
clouds, causing a burst of star formation at a rate beyond that in regular
spiral (S) galaxies \citep{2005ASSL..329..143S}, and the dust content of the
ISM is known to increase with the specific star formation rate
\citep{2010MNRAS.403.1894D, 2020MNRAS.496.3668D}.  Perhaps together with dust
from at least one S galaxy progenitor, the merger-induced dusty starbursts set
the dust-poor and dust-rich S0 galaxies apart.  Shocks can also reduce or
remove the orbital angular momentum of gas clouds, causing the gas to fall toward
the centre of the merger product and trigger a starburst \citep{1988ApJ...325...74S}.  The creation of tidally-induced bars may
also torque the gas clouds and drive them (someway) inward, where further
star formation may occur.\footnote{Nuclear discs and nuclear bars \citep[aka secondary bars,
  e.g.,][]{1990ApJ...363..391P} may funnel gas into AGN
\citep{1989Natur.338...45S, 1997ApJS..110..299M}.  The abundance of nuclear
stellar discs in S0 galaxies \citep[e.g.][]{1994AJ....108.1567J,
  2001AJ....121.2431R, 2007ApJ...665.1084B} may be a sign/means, even if a
relic, of AGN fuelling.  Counter-rotating nuclear discs are, of course, also a
sign of past accretion events.}   All this is to say that the environment matters.
However, a galaxy's environment has also taken something of a back seat to the
assumed dominance of AGN feedback in shaping the coevolution of galaxies and
black holes.

The observation that quasars reside in lower density regions than radio
galaxies \citep{2011A&A...535A..21L} 
meshes with the notion that once galaxies are in a (cold gas)-stripping
environment, their growth shuts down \citep{2008ApJ...673..715C}.  
Radio mode feedback (from a `Benson Burner') essentially maintains the status
quo in galaxies with `radio mode' AGN.  For these galaxies, subsequent evolution across the 
black hole scaling diagrams no longer
occurs through AGN accretion or star formation but via 
mergers \citep{Graham:Sahu:22a, Graham:Sahu:22b}, building bigger E galaxies
and brightest cluster galaxies (BCG). 

It might be tempting to speculate that the trend observed for the dust to reside in the more
massive S0 galaxies reflects their ability to (gravitationally) hold on to it,
while the lower-mass S0 galaxies cannot.  However, although feasible, 
the equally lower-mass spiral 
galaxies can hold onto their dust, with many maintaining a
Seyfert at their centre.  Therefore, there must be more afoot than simply the
stellar mass.  The
environment may come into play, with many lower-mass S0 galaxies ram-pressure stripped of gas due
to their residence in, or passage into, a hot X-ray gas cloud.  Although,
contradicting matters is that some visibly 
dust-poor S0 galaxies contain hydrogen gas \citep[e.g.,][and references
  therein]{2017A&A...605A..74K}.  Spiral galaxies are, however, known to
have a preference for existing in the field and group environment.

To probe this plausible but speculative idea --- of a (dust and
gas)-stripped galaxy --- the broad environment of the lenticular galaxies was
tracked down, specifically, if they are isolated or belong to a group or a
cluster.  This information has been recorded in the Appendix but, on its own, 
appears not to offer clarity.  This may, in part, be due to the randomness of
mergers in small groups or whether the group has entered a cluster's X-ray
halo --- which has not been explored.

\subsubsection{Faded S galaxies} 

There is some
suggestion in Fig.~\ref{Fig-6panel} that the more massive S0 galaxies ($M_{\rm *,gal}
\gtrsim 7\times10^{10}$ M$_\odot$; $M_{\rm *,sph} \gtrsim 2.5\times10^{10}$
M$_\odot$) retain their dusty gas, at least prior to the onset of a hot X-ray
halo which may heat the gas and destroy the dust.  If so, then perhaps the
five (less massive) dust-poor S0 galaxies (NGC: 1023; 3384; 4371; 4762; and 7332)
tracking the lower-end of the spiral galaxy $M_{\rm bh}$-$M_{\rm
  *,gal}$ relation\footnote{Four of the above five have
bulges to the right of the S galaxy $M_{\rm bh}$-$M_{\rm *,sph}$ relation
(Fig.~\ref{Fig-MMsph}).} were built by a wet
merger and should, in an evolutionary sense, be grouped with the dusty S0
galaxies.  Perhaps the dusty clue to their past has been swept away.  
Their proximity to the spiral merger NGC~2960 --- the pink hexagon
with the lowest mass in Fig.~\ref{Fig-MMgal} --- also adds credence to the 
notion (speculation) that these S0 galaxies might be faded S galaxies, or
faded S0 galaxies, 
`frozen-in' (into their resting place) in the $M_{\rm bh}$-$M_{\rm *,sph}$
diagram. 

The offset between the dusty S0 galaxies, the S galaxies, and the dust-poor S0
galaxies may blend well with the idea that S galaxies might fade into S0
galaxies once they lose their cold gas \citep{1980ApJ...237..692L,
  2002ApJ...577..651B, 2009MNRAS.394.1991B}.  However, the average $\sim$0.3
dex difference in the galaxy stellar mass, at fixed black hole mass, between
the S galaxy and the dust-poor S0 galaxy $M_{\rm bh}$-$M_{\rm *,gal}$
relations may pose too large a stumbling block for this scenario.  One
alternative is dramatic black hole growth from a circumnuclear gas disc with
relatively little global star formation, effectively pumping the S galaxies up onto the
dust-poor S0 galaxy relation in the $M_{\rm bh}$-$M_{\rm *,gal}$ diagram.

As noted above, rather than evolve from the S galaxy relation to the dust-poor S0 galaxy
relation, the tidal-stripping of gas may cause the S galaxies to freeze in the $M_{\rm
  bh}$-$M_{\rm *,gal}$ diagram and not undergo further movement. In contrast, 
those with an ongoing (cold gas)-supply continue the march to higher
stellar and black hole masses.  The dust-poor S0 galaxies with high $M_{\rm
  bh}/M_{\rm *,sph}$ ratios on the upper left-hand side of the mass-mass scaling
diagram may then be something of a relic population\footnote{NGC~7457 does not
  conform to this picture, unless it experienced something of a minor
  merger.}, frozen-in long ago and 
revealing how the (cold gas)-rich $M_{\rm bh}$-$M_{\rm *,gal}$ relation looked
in the past before it became today's dust-poor S0 galaxy relation.

\subsubsection{Non-exponential discs}

Simulations have shown that when discs are present, stars from infalling
satellite galaxies have a tendency to align with and heat the pre-existing
disc \citep[e.g.][]{2017Galax...5...44J}.  As the captured gas cools, modulo
any ongoing gravitational perturbations, it settles to the mid-plane (e.g.,
NGC~4233, NGC4710, NGC~5866) and forms a new generation of stars \citep{1980MNRAS.193..189F}.  This
captured material roughly follows an exponential light profile, given the
observed light profiles of disc galaxies. However, it 
may contribute to some anti-truncated discs by disproportionately adding
material to the pre-existing disc at either large or small radii, thereby
producing a double-exponential disc light profile \citep{2009ApJ...700.1896K}.  It is noted that other
mechanisms may also be capable of doing this, such as bars which can enhance
the, or arguably create an additional, inner disc \citep{2005ApJ...626..159B}.
Plus, the cluster environment appears to reduce the stellar disc scalelengths
\citep{2004ApJ...602..664G} and may cause disc truncations
\citep{1981A&A....95..105V}.  Given the smaller stellar masses of the
dust-poor S0 galaxies relative to the dust-rich S0 galaxies
(Fig.~\ref{Fig-MMgal}), they will naturally have smaller disc sizes.  It might
be insightful to check for disc truncations and
anti-truncations among the dust-poor and dust-rich S0 galaxies.  Such an 
investigation will be left for elsewhere.

\subsection{Blue ETGs and star formation} 

Studies of dust in ETGs \citep[e.g.][]{2012ApJ...748..123S} will
benefit from a division of not just E versus S0 galaxies but dust-poor versus
dust-rich S0 galaxies.
This division is likely related to the discovery of a population of blue spheroids/ETGs
\citep{2005MNRAS.363.1257E, 2007ApJ...657L..85D}, half of which have moderate to high
star-formation rates \citep{2009AJ....138..579K,
2009MNRAS.396..818S, 2018MNRAS.475..788M}. 
The low-mass blue ETGs have already been called out as major merger remnants
fading onto the red sequence \citep{2009AJ....138..579K}. 
As the star-formation in these
wet-merger-built systems ceases, their emission lines first disappear while
they remain blue before they become dusty red ETGs, like
NGC~1316.  
Therefore, it is not simply a matter of 
star-forming versus non-starforming ETGs, or 
blue versus red ETGs: both blue ETGs and dust-rich
red ETGs are expected to be offset from the dust-poor S0 galaxies that did not
experience the continuation of wet mergers
that built-up the dust-rich S0 galaxies.  

As shown here, a deeper insight into a galaxy's evolution can come from
knowledge of its morphological type (E, ES, S0), $M_{\rm bh}/M_{\rm *,sph}$
and $M_{\rm bh}/M_{\rm *,gal}$ ratio, and visible signs of dust.  For
instance, \citet{2009MNRAS.396..818S} found that 5 per cent of their ETG
sample is blue, reflecting current/recent star formation (in the absence of
substantial dust reddening).  By looking at the dust rather than the colour,
it is found here that roughly half of the S0 galaxies experienced a major wet
merger. This is a ten-fold increase in systems that experienced star formation
from a major wet merger, larger in part because it does not only capture
systems currently/recently undergoing star formation and in part because
hidden, dust-reddened star-forming systems are captured.  An additional reason
is evident when comparing with \citet{2006MNRAS.368..414D}, which rewrote the
notion that all ETGs are `red and dead'.  They found a one-third blue fraction
of ETGs.  The difference is because \citet{2006MNRAS.368..414D} used a sample
of galaxies extending $>$3 mag fainter than the L$^*$ or brighter sample of
\citet{2009MNRAS.396..818S}.  In descending to these fainter luminosities, it
is noted that although dwarf spiral galaxies are rare, some dwarf S0 galaxies
do contain weak spiral structures \citep[e.g.][]{2000A&A...358..845J,
  2002A&A...391..823B, 2003AJ....126.1787G}.  There is also a population of
blue compact dwarf (BCD) galaxies, comprised of an old underlying disc
coloured blue by accretion-induced star formation
\citep[e.g.][]{2008MNRAS.388L..10B, 2009A&A...501...75A, 2022ApJ...938...96J}.

\citet{2016ApJ...817...21S} introduced the notion of a blue and red sequence
in the $M_{\rm bh}$-$M_{\rm *,sph}$ and $M_{\rm bh}$-$M_{\rm *,gal}$ diagram,
with the LTGs defining a steep relation and the ensemble of ETGs defining a
near-linear relation.  However, it is now understood that the red sequence was
artificial.  With dusty S0, dust-poor S0, and E galaxies all following a steep
relation, one can add a third parameter (replacing morphology) to the $M_{\rm
  bh}$-$M_{\rm *,sph}$ diagram to show the star-forming galaxies, which
includes not only S galaxies but some of the younger S0 merger products like
NGC~5128.  A galaxy's star formation rate can be estimated in many ways. One
is from the blackbody glow at infrared wavelengths of the heated
dust. \citet{2016ApJ...830L..12T} used this approach, and 
interpreted the trend in the $M_{\rm bh}$-$M_{\rm *,gal}$ diagram as strong
evidence of AGN feedback, such that 
a low specific black hole mass, i.e., lower $M_{\rm bh}/M_{\rm                                                       
 *,}$ ratio, was less capable of quenching star formation
or, conversely, that higher $M_{\rm bh}/M_{\rm *,gal}$ ratios
in (some of) the ETGs were a signature of AGN suppression of star formation.
However, the interpretation in the current paper is that 
the systems with higher star formation rates primarily have these because of                                                     
 external refuelling events dictating their morphology
and shaping their distribution in the $M_{\rm bh}$-$M_{\rm *,gal}$ diagram.  

A variant of this theme for visualising the blue/red sequence of 
\citet{2016ApJ...817...21S} was used by \citet[][their
  Fig.~3]{2020ApJ...898...83D}, who showed the NUV-[3.6] colour as the third
parameter.  \citet{2014MNRAS.444.3408Y} had previously used the NUV-K colour
for the ATLAS$^{\rm 3D}$ galaxies to explore how it relates to their cold gas
content.  While \citet{2014MNRAS.444.3408Y} concluded that S0 galaxies are
quenched S galaxies that underwent bulge growth due to cold gas accretion and
modest star formation in some, it seems plausible that the dusty S0 galaxies
studied by \citet{2014MNRAS.444.3408Y} are instead the result of an
S0-building wet merger involving at least one S galaxy.

Counter-rotating gas discs, counter-rotating stellar discs, and polar discs
\citep[e.g.,][]{2004AJ....127.2641S} have long been recognised as signatures
of accretion or mergers \citep[e.g.,][]{1992ApJ...401L..79B, 1996MNRAS.283..543K}.
Past events can also reveal themselves as chemically distinct features
\citep[e.g.,][]{2000AJ....120..741S}.  Regarding the counter-rotating inner
dust disc of NGC~3032 (not in our sample), \citet{2008ApJ...676..317Y} suggest
that the ``molecular gas was captured through cold accretion from the
intergalactic medium or in an interaction or a minor merger with a gas-rich
neighbor''.  They add that, ``Perhaps it is even a remnant of a major merger
which formed the present galaxy. The high degree of regularity in the gas
kinematics and stellar morphology suggests that this event did not occur
recently.''  Several galaxies in our sample appear to fit this bill --- albeit
with co-rotating discs, which are a more probable occurrence --- such as 
NGC~4459, a likely merger with a high bulge-to-total stellar mass ratio of $\sim$0.6, 
and NGC~4526, which was also studied by \citet{2008ApJ...676..317Y}.

\subsection{ALMA's window}

Long-baseline submillimetre-to-radio interferometry currently enables better
spatial resolution than single aperture mirrors, yielding measurements of
smaller and/or more distant black holes \citep[e.g.][]{2019A&A...623A..79C,
  2021MNRAS.504.4123N, 2021MNRAS.503.5984S}.  Such measurements do, however,
require specific conditions, such as a cool, stable gas disc or ring that does
not dominate the mass budget within the sampled volume around the black hole.
This has proved most successful regarding maser emission, with some two dozen
and counting black hole masses measured this way \citep[e.g.][and references
  therein]{1995Natur.373..127M, 2020MNRAS.498.1609K}.  Galaxies with such
favourable conditions\footnote{While ionised gas clouds may not be dynamically
  relaxed, cold molecular gas which has settled into a disc is likely to
  follow Keplerian orbits.}  are overwhelmingly spiral galaxies, with a few
dusty S0 mergers (e.g., NGC~1194 and NGC~2960) also in the mix.  The spiral
galaxies define steep $M_{\rm bh}$-$M_{\rm *}$ relations \citep[][plus Fig.~4
  and 6]{2016ApJ...817...21S, 2019ApJ...873...85D}, with only the most massive
overlapping with, and thus appearing to conform with, the near-linear $M_{\rm
  bh}$-$M_{\rm *,sph}$ relation dominated by restricted samples of ETGs (no
`over-massive' black holes, no obvious dusty mergers, not many low-mass
dust-poor S0s which act to steepen the slope. This does not mean that the bulk
of the maser sample does not follow black hole scaling relations but instead
reveals the inadequacy and limitation of the old near-linear relation.

The tendency for the more massive S0 galaxies to contain visible signs of dust
(Fig.~\ref{Fig-6panel}) meshes well with the observation that massive S0
galaxies tend to have carbon-monoxide (CO) discs \citep{2013MNRAS.432.1796A}.
Given the Atacama Large Millimeter/submillimeter Array's (ALMA's) ability to
detect the emission from nuclear CO discs, ALMA's window into the black hole
scaling diagrams may not offer a full view of the black hole landscape.  For
instance, if the occurrence of nuclear molecular discs predominantly coincides
with the presence of visible dust and cold gas on large scales, then observing
programmes to improve the black hole scaling relations may be limited to dusty
galaxies, thereby offering a somewhat limited view and potentially curtailing
the interpretation of the data and the scaling relations.  For example, the
dust-poor S0 galaxies and the (merger-built) elliptical galaxies, which occupy
the lower-left and upper-right, respectively, of the distribution in the
$M_{\rm bh}$-$M_{\rm *,gal}$ diagram (Fig.~\ref{Fig-M-S0-gal} and
\ref{Fig-MMsph}), might be relatively sparsely sampled relative to the subset
of dusty S0 and S galaxies.  That is, observing campaigns with ALMA might be
restricted to sampling a limited/biased distribution of systems across the
Jeans-Lundmark-Hubble galaxy sequence \citep[][and references
  therein]{1919pcsd.book.....J, 1925MNRAS..85..865L, 1926ApJ....64..321H,
  1927UGC..........1L, 1928asco.book.....J, 2019MNRAS.487.4995G}.
Although, it remains to be seen how many types of galaxy ALMA will be able to
properly sample.  

Moreover, coupled with an awareness of the morphology-dependent black hole
scaling relations, submillimetre-to-radio interferometers are not only
refining previous black hole mass measurements and bolstering statistics with
increased numbers, but they should result in further insight into the
unfolding story of coevolution via mergers, fuelling (star formation and AGN),
and galactic speciation. 
Studies of CO and other molecules, such as hydrogen cyanide (HCN), in the discs
of high-$z$ galaxies offer 
exciting pathways to witness further the growth of black holes and the
metamorphosis of their host galaxies \citep[e.g.][and references
  therein]{2007PhDT........22R, 2017MNRAS.468.4205K,  2022ApJ...941..106F, 
2022A&A...665A.107T}.  Awareness of the morphology-dependent black hole
scaling relations will be necessary for the comparison with local galaxies and
interpretation.

\section{Future work}\label{Sec_Future}

The discovery of interlaced $M_{\rm bh}$-$M_{\rm *,sph}$ relations (from
dust-poor S0 galaxies, S galaxies, dusty S0 galaxies to E galaxies) invites 
further quantitative probing.  The following list mentions
many potential research endeavours which could be pursued to help shore up, refine, or modify
the new picture of galaxy/black hole coevolution involving `punctuated
equilibrium' jumps through the different galaxy types.
There may additionally be a kind of `gradualism', or `secular evolution',
along the relations for the different morphological types, e.g., Sd toward Sa
and $M_{\rm bh}/M_{\rm *,sph}$ growth among the dusty S0 galaxies if the
accreted gas from the merger grows the black hole at a faster rate than star
formation \citep{2008MNRAS.387.1163C, 2009MNRAS.396..423B,
  2012MNRAS.420.2662D, 2012ApJ...755..146S}.

The following list is also intended to help indicate the breadth of galaxy research
pertaining to, and stemming from, the black hole scaling relations.

\begin{itemize}

\item  The location of a disc galaxy in the $M_{\rm bh}$-$M_{\rm *,sph}$ scaling diagrams
could be explored as a function of the (cold gas)-to-stellar mass ratio,
normalised to the stellar mass: atomic H{\footnotesize I} or molecular CO and H$_2$ gas. 

\item  Consider the specific dust mass as a third parameter. 
Recognition of the dusty S0 galaxies as something of an intermediate-stellar-mass
  population between dust-poor S0 galaxies and E galaxies may bring crucial
  insight for studies trying to connect dust mass with other properties of
  ETGs \citep{2012ApJ...748..123S}.  

\item  Check the  Polycyclic Aromatic Hydrocarbon (PAH) emission 
\citep{2008ApJ...684..270K, 2012AJ....143...49W, 2019PASJ...71...25K} versus
  location in the $M_{\rm bh}$-$M_{\rm *,sph}$ diagram.

\item Check on the presence of hot X-ray gas. In particular, are the low-mass S0s
located within such an environment, acting to freeze them in time, modulo
cluster-related tidal
effects snapping at their heels and eating away at them? 

\item  The
dominant gas-removal mechanism for many individual galaxies could be explored, as was
carefully done for the dwarf galaxy DDO~113 by \citet{2020MNRAS.492.1713G}.

\item The disc-to-bulge stellar mass ratio, disc scalelengths, and
  bulge-to-disc size ratios could  be
explored regarding cluster-induced truncations
\citep{2004ApJ...602..664G}. 

\item The presence of truncated or anti-truncated stellar discs could be
explored to see if they predominantly occur in a particular part of the diagram.

\item
The star formation history could be examined
\citep[e.g.][]{2007IAUS..241..175C, 2011ApJ...739....5W}, adding this information as a (coloured)
third parameter in the scaling diagram.   
Another option includes showing the amount of H$\alpha$ luminosity from star
formation. 

\item 
Beyond the $M_{\rm bh}$-$M_{\rm *,gal}$, $M_{\rm bh}$-$M_{\rm *,sph}$, and
$M_{\rm bh}$-$R_{\rm e,sph}$ diagrams seen here, it may be insightful to
explore the location of visibly dusty (wet merger)-built S0 galaxies and dust-poor S0
galaxies in the $M_{\rm bh}$-$\sigma$, $M_{\rm bh}$-$n$, and $M_{\rm
  bh}$-$\rho$ diagrams \citep{2000ASPC..197..221M, 2007ApJ...655...77G, 2016ApJ...818...47S,
  2022ApJ...927...67S}. 

\item 
Explore if the mass scaling relation between nuclear star clusters and their
host spheroid, discovered two decades ago \citep{2003ApJ...582L..79B,
  2003AJ....125.2936G}, also displays a dependence on galaxy morphology.
This may reveal insight into their coevolution. 

\item 
Investigate whether the (nuclear star cluster mass)-(black hole mass)
relation, first presented at a conference in China in August of 2014
\citep{2016IAUS..312..269G} and more recently refined by 
\citet{2020MNRAS.492.3263G}, 
has any dependence on the host galaxy properties.

\item Pursue the interaction and accretion
history of all the spiral galaxies in the sample. 

\item  Investigate the abundance of bars (and nuclear bars) in dust-poor and
  dust-rich S0 galaxies, and S galaxies, and also along their $M_{\rm bh}$-$M_{\rm
    *,sph}$ sequences. 

\item The S0 galaxies built from gas-rich mergers (of, say, an S0 and an S
  galaxy or two S galaxies) may display evidence of this merger in their
  globular cluster systems (GCSs).  This is evident in the Milky Way, where
  detailed information has revealed at least three past mergers
  \citep{2019MNRAS.486.3180K}, and it may be evident in the dusty S0 galaxies
  as bimodal and trimodal colour distributions of globular clusters
  \citep{1997AJ....113.1652F, 1999IAUS..186..173Z, 2006ARA&A..44..193B}.  In
  contrast, the dust-poor S0 galaxies may have a more simple GCS. Building on
  \citet{2013MNRAS.433..235P} and \citet{2022ApJ...941...53G}, a programme to
  measure the colour histograms of the GCSs of galaxies with directly measured
  black masses may be insightful, or even comparing the histograms in just the
  dust-poor versus dust-rich S0 galaxies could be telling.

\item 
Improved predictions can now be made about where intermediate-mass black holes
(IMBHs) may reside.  Given the near-absence of dwarf S galaxies, the
super-quadratic scaling relation for the dust-poor S0 galaxies appears to
offer the most promise for predicting the masses of IMBHs in dwarf galaxies,
perhaps even providing a direct bridge to the lower end of the IMBH range
$10^2 < M_{\rm bh}/M_\odot < 10^5$ \citep{2022A&A...659A..84A}.

\item Check if simulations focussed on S galaxies reproduce the steep $M_{\rm
  bh}$-$M_{\rm *,gal}$ relation and the steep $M_{\rm bh}$-$V_{\rm rot}$ and
  $M_{\rm bh}$-$M_{\rm dark-matter}$ relations \citep[][and references
  therein]{2019ApJ...877...64D}.

\end{itemize}

\section*{Acknowledgements}

The author is grateful for discussions with Anita Pappas and David Brown,
Billanook College. 
Part of this research was conducted within the Australian Research Council's
  Centre of Excellence for Gravitational Wave Discovery (OzGrav) through
  project number CE170100004.
This work has used the NASA/IPAC Infrared Science Archive (IRSA) 
and the NASA/IPAC Extragalactic Database (NED),
funded by NASA and operated by the California Institute of Technology. 
Based on observations made with the NASA/ESA Hubble Space Telescope,
obtained from the Mikulski Archive for Space
Telescopes and the Hubble Legacy Archive. 
This research has also used the NASA/SAO Astrophysics Data System Bibliographic
Services and the {\sc Rstan} package available at \url{https://mc-stan.org/}.

\section{Data Availability}

The imaging data underlying this article are available in the NASA/IPAC
Infrared Science Archive.  The derived spheroid and galaxy stellar masses are 
tabulated in \citet{Graham:Sahu:22a}.

\bibliographystyle{mnras}
\bibliography{Paper-BH-mass3}{}

\appendix

\section{Environment} 
\label{Apdx1}

\subsection{Cluster, group, field environment of the S0 galaxies}

The environment, whether field, group or cluster (or immersion within a hot
X-ray-emitting gas cloud), has long been known to play a crucial role in the
evolution of galaxies through its influence on the presence of cold gas.  Despite
this, the black hole scaling relations are yet to recognise and explicitly
incorporate an environmental parameter.  In this paper, we have seen that, for
a given $M_{\rm bh}$, S galaxies have bigger spheroid stellar masses than
dust-poor S0 galaxies, and dust-rich S0 galaxies, known to be built by major
wet mergers, have even larger spheroid stellar masses for the same black hole mass.  These
(galaxy morphology)-dependent $M_{\rm bh}$-$M_{\rm *,sph}$ relations imply a
gas dependence in the scaling diagrams that has yet to receive much
attention and may, in part, be controlled by the environment
\citep[e.g.][]{2012ApJ...745...13S, 2018MNRAS.480.1022Y}.

Indeed, some galaxy environments are less conducive to wet mergers than others.
Clusters, for example, may `dry out' galaxies, removing their gas.  Furthermore,
the high speeds of galaxies in clusters reduce the chances of galaxy
collisions.  Collectively, this could leave a population of dust-poor S0 galaxies unlikely
to acquire new gas and evolve. Unless the BCG assimilates them, they would, in
a sense, be frozen in time, albeit containing an ageing stellar population.

Hot haloes of X-ray emitting gas have likely halted star formation and quasar activity
in elliptical galaxies \citep[][and references therein]{2003ApJ...599...38B,
  2009ApJ...696..891H, Graham:Sahu:22a}.
The NGC~5813 subgroup associated with the N5846 Group in the sprawling Virgo
cluster is one such example, with a `Benson
Burner' at the heart of the S0 galaxy 
NGC~5813 keeping the gas hot \citep{2015ApJ...805..112R}.  
This mechanism  may also contribute to the 
tendency for some larger spiral galaxies to be deficient in atomic hydrogen  
\citep[e.g.][]{1991ApJ...374..103V}. 
In addition to thermal evaporation \citep{1977Natur.266..501C}, 
cluster-sized X-ray haloes can lead to ram-pressure stripping of 
cold H{\footnotesize I} gas \citep[][]{1972ApJ...176....1G, 1973MNRAS.165..231D,
1985ApJ...292..404G, 2010AJ....140.1814Y, 2021ApJ...915...70W}. 
Group-sized X-ray bright haloes can also remove gas from a
galaxy \citep{1979ApJ...234L..27F} and thus
strangle the supply of cold gas, which 
may have otherwise arisen from the cooling of hot gas \citep{2008ApJ...672L.103K, 2015Natur.521..192P}. 
Cold-gas removal mechanisms appear to also operate in compact groups 
and some loose groups \citep[e.g.][]{2001A&A...377..812V, 2005JApA...26...71O,
2007MNRAS.378..137S, 2022MNRAS.515.5877K}. 
The affected galaxies are observed to have reduced H{\footnotesize I} content
and reduced H{\footnotesize I} disc sizes.  Reduced H$\alpha$
sizes have also been observed \citep{2020MNRAS.496.3841V}, and 
reduced stellar disc sizes in galaxy clusters were discovered two decades ago  
\citep{2004ApJ...602..664G}, 
likely also a result of gravitational tides \citep{Roche:1850} from repeated fleeting encounters
with each other 
\citep{1990AJ.....99.1740H, 1998ApJ...495..139M, 2022ApJ...927...66W}, thereby
building up the intracluster light.  

Table~\ref{Table-env} provides the S0 galaxy environments gleaned from the
literature.
Much of the group membership can be found in \citet{1993A&AS..100...47G},
later refined by \citep{2004MNRAS.350.1511O, 2011MNRAS.412.2498M}.  
No apparent trends have been found between the dust-rich and dust-poor S0 galaxies.

\begin{table*}
\centering
\caption{Environment of the S0 and ES,b galaxies}\label{Table-env}
\begin{tabular}{lcl} 
\hline
Galaxy    & Dust &  Environment  \\
\hline 
\multicolumn{3}{c}{19 largely non-dusty (excluding small nuclear dust rings/discs) S0 galaxies} \\
NGC~1023  &  N  &      NGC~1023 Group (N=5) \\ 
NGC~1332  &  n  &  BGG in NGC~1332 Group (N=22) in Eridanus Cl.\ assoc.\ with Fornax Cl.\\ 
NGC~1374  &  N  &      Fornax Cl.\ NS Clump (N=47) \citep{2019AandA...627A.136I} \\ 
NGC~2549  &  N  &      Isolated pair \\ 
NGC~2778  &  N  &      LGG 171. NGC~2778 Group  (N=3) \\ 
NGC~3245  & n/y &      in NGC~3256 Group (N=6) within Leo II Groups \\  
NGC~3384  &  N  &      in Leo I Group. NGC~3379 Group (N=27) \\  
NGC~4339  &  N  &      M49 Group, aka NGC~4472 Group or Virgo~B Cl.\ (N=100+) \\ 
NGC~4342  &  N  &      NGC~4342 Group (N=5), within M49 Group \\ 
NGC~4350  &  n  &      pair with NGC~4340 within M49 Group \\  
NGC~4371  &  n  &      LGG 292 (N=59) within M49 Group \\ 
NGC~4434  &  N  &      LGG 292 (N=59) within M49 Group \\ 
NGC~4564  &  N  &      M49 Group (N=100+) \\ 
NGC~4578  &  N  &      NGC~4568 Group (N=6) \\  
NGC~4742  &  N  &      NGC~4699 Group (N=15) within NGC~4697 Group (N=37) \\ 
NGC~4762  &  N  &      poss.\ past merger in binary with N4754. LGG 292 (N=59) within M49 Group \\ 
NGC~5845  &  n  &      LGG 392 (N=6) within NGC~5846 Group (N=74) \\ 
NGC~7332  &  N  &      isolated binary with NGC~7339 \\ 
NGC~7457  &  N  &      NGC~7457 Group (N=2) \\ 
\multicolumn{3}{c}{21 dusty S0 galaxies} \\
NGC~0404  &  Y  &      isolated galaxy \\ 
NGC~0524  &  Y  &  BGG in NGC~0524 Group (N=16), assoc.\ with NGC~0488 Group (N=17) \\ 
NGC~1194  &  Y  &      pair \\ 
NGC~1316  &  Y  &  BCG in Fornax Cl.\ (N=100+) \\ 
NGC~2787  &  y  &      isolated, but poss.\ merger in NGC~2787 Group (N=2) \\  
NGC~2974  &  Y  &      isolated NGC~2974 Group (N=5) \\
NGC~3115  &  Y  &      isolated NGC~3115 Group (N=5) \\  
NGC~3489  &  Y  &      isolated, but within the Leo I Group (N=27)\\ 
NGC~3665  &  Y  &  BGG in NGC~3665 Group (N=11--16), in M49 Group (N=100+)\\ 
NGC~3998  &  y  &      LGG 241 (N=8) within NGC~3992 (aka M109) Group (N=72) \\ 
NGC~4026  &  y  &      M109 Group (N=72) within Ursa Major Cl. \\ 
NGC~4429  &  Y  &      poss.\ old merger, in M49 Group (N=100+) \\ 
NGC~4459  &  Y  &      M49 Group (N=100+) \\ 
NGC~4526  &  Y  &      M49 Group (N=100+) \\ 
NGC~4594  &  Y  &  BGG in NGC~4594 Group (N=11) within Virgo B Cl. \\
NGC~4596  &  Y  &      NGC~4535 Group (N=23) within the Virgo Cl. \\ 
NGC~5018  &  Y  &  BGG and old merger in NGC~5018 Group (N=9) assoc.\ with NGC~5044 Group (N=52) \\ 
NGC~5128  &  Y  &  BGG Cen-A merger in NGC~5128 Group (N=15) \\ 
NGC~5252  &  Y  &      NGC5252 Group (N=6) \citep{1996AnA...306...39F, 2017MNRAS.470.2982L} \\ 
NGC~5813  &  y  &      NGC~5846 Group (N=74) within the Virgo III Groups \\ 
NGC~6861  &  Y  &      LGG 430 (N=11). 2nd brightest in the NGC~6868 Group (N=19) aka Telescopium Group \\ 
\hline
\end{tabular}

Column~1: Galaxy name. 
Column~2: Is there dust visible in the HST optical image: 
(Y)es plenty, 
(y)es but not much, 
only a small (n)uclear dust disc/ring, or 
(N)othing obvious. 
Column~3: Environment.  Common knowledge supplemented primarily by the 
Lyon Groups of Galaxies (LGG) catalogue \citep{1993A&AS..100...47G} and the 
group catalogue of \citet{2011MNRAS.412.2498M}, accessed via the Extragalactic
Distance Database 
\citep[\url{http://edd.ifa.hawaii.edu/}][]{2009AJ....138..323T}. 

\end{table*}

\bsp    
\label{lastpage}
\end{document}